\documentclass[twocolumn,nofootinbib]{revtex4-1}
\usepackage{amssymb}
\usepackage{amsmath}
\usepackage{latexsym}
\usepackage{graphicx}
\usepackage{slashed}
\usepackage{pifont}
\usepackage{color}
\usepackage[usenames,dvipsnames]{xcolor}

\newcommand{\req}[1]{Eq.\,(\ref{#1})}
\newcommand{\beqn}{\begin{equation}}
\newcommand{\eeqn}{\end{equation}}

\newcommand{\Tr}{\mathrm{Tr}}

\begin{document}

\title{Robust characteristics of nongaussian fluctuations from the NJL model}
\author{Jiunn-Wei Chen}
\email{jwc@phys.ntu.edu.tw}
\affiliation{Department of Physics and Center for Theoretical Sciences,
National Taiwan University, Taipei, 10617 Taiwan}
\affiliation{Leung Center for Cosmology and Particle Astrophysics (LeCosPA),
National Taiwan University, Taipei, 10617 Taiwan}
\affiliation{Helmholtz-Institut f\"{u}r Strahlen- und Kernphysik and Bethe Center
for Theoretical Physics, Universit\"{a}t Bonn, D-53115 Bonn, Germany}
\author{Jian Deng}
\email{jdeng@sdu.edu.cn}
\affiliation{Key Laboratory of Particle Physics and Particle Irradiation (MOE),
School of Physics, Shandong University, Jinan 250100, China}
\author{Hiroaki Kohyama}
\email{kohyama.hiroaki@gmail.com}
\affiliation{Department of Physics and Center for Theoretical Sciences,
National Taiwan University, Taipei, 10617 Taiwan}
\author{Lance Labun}\email{labun@utexas.edu}
\affiliation{Department of Physics, University of Texas, Austin}

\date{7 September, 2015}

\begin{abstract}
We evaluate the third- and fourth-order baryon, charge and strangeness susceptibilities near a chiral critical point using the Nambu-Jona-Lasinio model.  We identify robust qualitative behaviours of the susceptibilities along hypothetical freeze-out lines that agree with previous model studies.  Quantitatively, baryon number fluctuations are the largest in magnitude and thus offer the strongest signal when freeze-out occurs farther away from a critical point.  Charge and strangeness susceptibilities also diverge at a critical point, but the area where the divergence dominates is smaller, meaning freeze-out must occur closer to a critical point for a signal to be visible in these observables.  In case of strangeness, this is attributable to the relatively large strange quark mass.  Plotting the third- and fourth-order susceptibilities against each other along the freeze-out line exhibits clearly their non-montonicity and robust features.
\end{abstract}

\maketitle



\section{Introduction}

Heavy ion collision experiments and lattice simulations are probing the
phase diagram of QCD matter to help understand the chiral and deconfinement
phase transitions \cite{QCDphases}. These experiments have shown that at low
density, the transition is a continuous cross-over at $T\sim 165$\thinspace
MeV \cite{Gupta:2011wh}. At high density, models lead to the expectation
that the transition is first order. This structure would be verified by
locating a critical end point in chemical potential--temperature plane where
the first order line begins. The heavy ion collision experiments yield
statistical observables of QCD matter, including proton number and electric
charge fluctuations \cite{Aggarwal:2010wy,Adamczyk:2013dal},and lattice
studies have developed a variety of approaches to searching for structure at 
$\mu _{B}\neq 0$ \cite%
{Fodor:2001au,Allton:2002zi,deForcrand:2010he,Gavai:2010zn,Li:2011ee,Endrodi:2011gv,Schmidt:2015kea}%
.

To interpret the experimental data in terms of the phase diagram, we need to
relate them to signatures derived from theory calculations. Since
conventional lattice methods are limited to low baryon density by the sign
problem, it is useful to complement the investigation with model studies in
which we can thoroughly study the phase structure at $\mu \neq 0$. We use
the models to identify robust signatures of the critical point, such as
consequences of the large correlation length, that would be manifest also
near a critical point in QCD. The correlation length becomes large near the
critical point (CP) because the CP is a second-order phase transition where
the mass term in the Landau effective potential for the order parameter $%
\sigma $ vanishes, $m_{\sigma }\sim \xi ^{-1}\rightarrow 0$. Therefore, at
the CP, the longest wavelength correlations (of order the system ``size'')
can be investigated using the partition function \cite{Stephanov:1998dy} 
\begin{equation}
\mathcal{Z}=\int \mathcal{D}\sigma e^{-\frac{1}{T}\int d^{3}x\Omega \lbrack
\sigma ]}  \label{partfnsigma}
\end{equation}%
with the Landau effective action 
\begin{equation}
\Omega \lbrack \sigma ]=\Omega _{0}+\frac{1}{2}(\vec{\nabla}\sigma )^{2}+%
\frac{m_{\sigma }^{2}}{2}\sigma ^{2}+\frac{\lambda _{3}}{3}\sigma ^{3}+\frac{%
\lambda _{4}}{4}\sigma ^{4}+...
\end{equation}%
$\Omega _{0}$ is the energy due to the vacuum expectation value (vev) of the
order parameter, $\sigma $ is the fluctuation around this vev, and for
fluctuations at wavelengths similar to the system size, we consider only the
zero momentum mode, so the kinetic term $(\vec{\nabla}\sigma )^{2}$ vanishes.

The divergent parts of the fluctuations associated with $\xi \rightarrow
\infty $\ at the CP (an infrared fixed point) are universal for systems
within the same universal class. The finite parts of the fluctuations are
model dependent which can be described by higher dimensional operators in
the effective action. Using Eq.\thinspace (\ref{partfnsigma}), statistical
observables such as baryon number fluctuations are related to fluctuations
of the order parameter \cite{Stephanov:1998dy}. For example, higher order,
non-Gaussian fluctuation moments are more sensitive to a critical point
because they diverge with a larger power of the correlation length, as shown
for the third and fourth order susceptibilities \cite%
{Stephanov:2008qz,Asakawa:2009aj,Stephanov:2011pb}.

In this approach, we hypothesize the presence of a critical point,
investigate the consequences, and compare them to data. It is important to
bear in mind that the conditions we identify for a critical point can only
be necessary conditions but not sufficient conditions. Also, there are many
assumptions in our attempt to compare with heavy ion data. For example, we
assume the fireball is near thermal equilibrium at freeze-out, though
critical slowing of dynamics would be important if the fireball approaches
the critical point chemical potential and temperature \cite{slowing}.
Additionally, there can be changes in expansion dynamics and interactions
that produce variations in particle spectra and acceptance independent of
critical phenomena, whose fluctuations must also be controlled for \cite%
{Koch:2008ia}. To claim evidence of a critical point in this context, we
should identify as many model-independent signatures as possible to
corroborate interpretation of the data in terms of a critical point.

In this work, we study the Nambu-Jona-Lasinio (NJL) model, which is a QCD
inspired field theoretic model with spontaneous breaking of Chiral symmetry 
\cite{NJL,Bernard:1987gw,Bernard:1987sg}. The model comprises distinct flavors of quarks that interact via
(effective) point-like 4-fermion operators. At low density and temperature,
the model exhibits a non-zero chiral condensate $\langle \bar{\psi}\psi
\rangle \neq 0$ \cite%
{Vogl:1991qt,Klevansky:1992qe,Hatsuda:1994pi,Buballa:2005rept,Huang:2004ik}.
The goal is to identify, in the NJL model results, characteristics of
fluctuation observables that are predicted to be model-independent by
analysis such as given in \cite{Stephanov:2008qz,Stephanov:2011pb}. Previous
work using the NJL model has focused on third moments \cite%
{Asakawa:2009aj}, or electric charge susceptibilities \cite%
{Skokov:2011rq}.  Numerical studies of the susceptibilities have
also used the Polyakov-loop improved NJL model \cite{Fu:2010ay} and Dyson-Schwinger approaches \cite{Xin:2014ela}.
Some of the model-independent characteristics have been
confirmed and cross-checked in the Ising \cite{Stephanov:2011pb} and
Gross-Neveu (GN) models \cite{Chen:2014ufa}.

There are two main goals in this work. The first goal is to make use of the
flavor dependence in the NJL model to compute the complete set of
susceptibilities on the phase diagram. The second goal is to check whether
some model-independent features obtained from a general effective potential
analysis then confirmed by the GN model \cite{Chen:2014ufa}\ will remain in
the NJL model, which belongs to the same universality class with QCD at the
CP. These features include: (1) The negative $\sigma $-kurtosis ($\langle
\sigma ^{4}\rangle -3\langle \sigma ^{2}\rangle ^{2}<0$) region occurs
almost entirely in the symmetric (normal) phase. (2) However, in addition to
the $\sigma $-kurtosis, there are more critical-mode correlators
contributing to the fourth-order baryon number susceptibility ($\partial
^{4}\ln Z/\partial \mu _{B}^{4}$) to determine its negative region. (3) The
peaks in non-Gaussian susceptibilities on a freeze out line obey an ordering
in temperature.

Among those features, we expect (1) to be robust, because they can be
understood from the shape of the effective potential around the critical
point \cite{Chen:2014ufa}. We also expect (2) to happen, that is, there are
other terms as important as the $\sigma $-kurtosis to determine the negative
region of ($\partial ^{4}\ln Z/\partial \mu _{B}^{4}$). This is based on a
general effective action analysis of Ref. \cite{Chen:2014ufa} to identify
terms with leading power divergence in $\xi $ near the CP. We will check it
numerically using the NJL model.

\section{Complete Set of Fluctuation Observables}

Fluctuations in conserved charges are important observables because they
can be obtained in principle from HIC experiments as well as lattice
simulations. Not all fluctuation moments can be measured in practice, and we
will discuss below the observationally most relevant subset. The fluctuation
moments are derivatives of the partition function with respect to the
chemical potentials of the conserved charges: 
\begin{equation}
\chi _{\alpha \beta }=\frac{\partial ^{2}\ln \mathcal{Z}}{\partial \mu
_{\alpha }\partial \mu _{\beta }},\quad \chi _{\alpha \beta \gamma }=\frac{%
\partial ^{3}\ln \mathcal{Z}}{\partial \mu _{\alpha }\partial \mu _{\beta
}\partial \mu _{\gamma }},...  \label{chin}
\end{equation}%
If $\alpha =\beta =....$ we shall also use $\chi _{\alpha }^{(n)}=\partial
^{n}\ln \mathcal{Z}/\partial \mu _{\alpha }^{n}$. Recall the first
derivative is the expected number, $\partial \ln \mathcal{Z}/\partial \mu
_{X}\equiv \langle N_{X}\rangle $, which is conserved and therefore
constrained by the initial conditions. For example, since protons and
neutrons comprise only up and down quarks and strangeness is conserved by
QCD reactions, $\langle N_{s}\rangle =0$. On the other hand, since the heavy
nuclei typically contain more neutrons than protons, the initial state has
an isospin asymmetry. 

At quark level, we have three independent chemical potentials $\mu _{u}$, $%
\mu _{d}$, and $\mu _{s}$\ associated with the conserved quark numbers for $%
u $, $d$, and $s$\ quarks, respectively. To compare to experiment,
we use the basis of conserved charges $\{B,Q,S\}$. The hadron-level
strangeness (large $S$) is defined so that $K^{+}$ has $S=+1$ in agreement
with the experimental notation and opposite the quark-level definition of
the strange charge (small $s$) where the strange quark has $s=-1$ and
antistrange quark has $s=+1$. We will use the $\{q,I,s\}$ basis of charges,
where $2\mu _{q}=\mu _{u}+\mu _{d}$ and $\mu _{I}=\mu _{u}-\mu _{d}$. Since
the isospin potential $\mu _{I}$ is also measurable at hadron-level, we can
use observations to constrain the chemical potentials, and so also the set
of susceptibilities we need to evaluate.

The chemical potentials $\mu _{B},\mu _{Q},\mu _{S},\mu _{I}$ at chemical
freeze-out are determined by fitting the observed particles yields to a
statistical model of hadronization in the fireball. Limited acceptance and
experimental effects can mean the fit values of the chemical potentials
differ from the theoretical expectation. The data indicate that when
collision energy per nucleon pair in the center-of-mass frame ($\sqrt{s_{NN}}$) is 7.7, 39, 200
GeV, $\mu _{B}/3$ is $\sim $130, 30, 10 GeV and $\mu _{S}$ is $\sim $90, 20,
5, respectively \cite{Das:2014oca}. Using Eq.\thinspace (\ref{quark2hadron})
and isospin symmetry $\mu _{u}\simeq \mu _{d}$, this implies $\mu _{s}$ is $%
\sim $40, 10, 5 GeV, even though theoretically $\langle N_{s}\rangle =0$ in
the initial state. However, given $\mu _{s}\lesssim $40 MeV which is smaller
than the lower range on the strange quark (current) mass $m_{s}\gtrsim 90$
MeV, it is a good approximation to set $\mu _{s}=0$ in our model
calculations: dependence of the phase diagram on $\mu _{s}$ is weak as long
as for $\mu _{s}\leq m_{s}$.

It is worth noting here that one piece of evidence for non-equilibrium at
chemical freeze-out is that particle yield fits are improved by including a
strange quark quenching factor $\gamma_s$, which observationally is $\leq 1$%
. This smaller-than-equilibrium abundance of both $s$ and $\bar s$ can be
understood considering that the phase space for producing $s,\bar s$ quarks
is smaller due to their moderately large mass $m_s\gtrsim 90$ MeV compared
to the temperature $T\lesssim 160$ MeV. That is, a substantially smaller
fraction of gluons (or $q\bar q$ pairs) collide with center-of-mass energy $%
\geq 2m_s>180$ MeV, which suppresses the reaction rate for production in
comparison to the lifetime of the fireball \cite{Petran:2013lja}. Although
it impacts the absolute abundance of measured strange particles, it is
linearly independent of $\mu_s$ that controls net strangeness, i.e. the
number of strange minus the number of antistrange quarks $N_s-N_{\bar s}$.
With $\mu_s=0$, there is no violation of strange quark number independent
the value of $\gamma_s$.

Additionally, analysis of the charged pion ratio $N(\pi ^{-})/N(\pi ^{+})$
has consistently concluded that $\mu _{I}=0$ at high beam energies, though
$\mu_I$ can differ from zero at the lowest collision energy $\sqrt{s_{NN}}=10$ GeV 
\cite{Cleymans:2004pp}. This variation as a function of $\sqrt{s_{NN}}$ can be
understood by the total particle multiplicity becoming much higher for
larger $\sqrt{s_{NN}}$. We have checked quantitatively in the NJL model, that the
phase diagram is weakly dependent on $\mu _{I}$ as long as $\mu _{I}\ll
m_{\pi }/2\simeq 65$ MeV (this is also found in \cite{Xia:2013caa}). 
Therefore, in our work we shall for simplicity set $\mu _{I}=0$ for all $\mu _{B}$.

With $\mu _{s}=\mu _{I}=0$, a complete set of nontrivial susceptibilities up
to fourth order is: 
\begin{align*}
& \chi _{qq},\chi _{II},\chi _{ss}, \\
& \chi _{q}^{(3)},\chi _{qss},\chi _{qII}, \\
& \chi _{q}^{(4)},\chi _{qqII},\chi _{qqss},\chi _{I}^{(4)},\chi _{IIss},\chi
_{s}^{(4)}
\end{align*}%
These are easy to count in the $\{q,I,s\}$ basis, because $\mu _{I}=\mu
_{s}=0$ implies that odd-order derivatives of $\mu _{I}$ vanish identically.
When we choose another basis, there can only be 3, 3, and 6 independent
susceptibilities at 2nd, 3rd and 4th order respectively. These are plotted
in Figs. \ref{fig:chi2-qIs}, \ref{fig:chi3-qIs}, \ref{fig:chi4-qIs-1} and %
\ref{fig:chi4-qIs-2}.

While the long-wavelength correlations are dominated by the iso-scalar
critical mode $\sigma $ and the isospin chemical potential is
\textquotedblleft small\textquotedblright\ (as defined below) this relation
shows that charge fluctuations diverge like the baryon fluctuations near the
critical point \cite{Hatta:2003wn}. As explained in the next section and
Appendix \ref{app:diagrams}, the second-order isospin and strange
susceptibilities do not have a singularity at the critical point, even in
the presence of flavor mixing. By direct computation of $\chi _{Q}$ and $%
\chi _{B}$ in the NJL model below, we will verify this fact. On the other
hand, because the correlation length in the fireball may peak at $\sim 3$
fm, only about twice the thermal correlation length (and much smaller than
infinity), the difference between the singular iso-scalar critical mode
contribution and the smooth model-dependent contributions to correlations
may be less dramatic.

To relate the $B,Q,S$ susceptibilities to the derivatives with respect to
quark susceptibilities we need to rewrite partition function in terms of the 
$B,Q,S$ charges. The grand canonical partition function can be expressed as 
\begin{equation}
\mathcal{Z}=\mathrm{Tr}e^{-\beta (\hat{H}-\mu _{X}\hat{N}_{X})}
\label{GCpartnfn}
\end{equation}%
where $\hat{N}_{X}$ is the number operator, i.e. the time component of the
conserved current. This implies that the quark chemical potentials $\mu
_{q},\mu _{I},\mu _{s}$ are related to the chemical potentials associated
with observable numbers $\mu _{B},\mu _{Q},\mu _{S}$ using the linear
relations between the particle numbers. Starting from the definitions 
\begin{equation*}
N_{q}\equiv N_{u}+N_{d},\quad N_{I}\equiv \frac{1}{2}(N_{u}-N_{d})\,,
\end{equation*}%
we have 
\begin{align}
N_{B}& =\frac{1}{3}(N_{q}+N_{s})=\frac{1}{3}(N_{u}+N_{d}+N_{s}) \\
N_{Q}& =\frac{1}{6}N_{q}+N_{I}-\frac{1}{3}N_{s}=\frac{1}{3}%
(2N_{u}-N_{d}-N_{s}) \\
N_{S}& =-N_{s}
\end{align}%
These relations lead to 
\begin{subequations}
\label{quark2hadron}
\begin{align}
\mu _{B}& =3\mu _{q}-\frac{1}{2}\mu _{I} \\
\mu _{Q}& =\mu _{I} \\
\mu _{S}& =\mu _{q}-\frac{1}{2}\mu _{I}-\mu _{s}
\end{align}%
and conversely 
\end{subequations}
\begin{subequations}
\label{hadron2quark}
\begin{align}
\mu _{q}& =\frac{1}{3}\mu _{B}+\frac{1}{6}\mu _{Q} \\
\mu _{I}& =\mu _{Q} \\
\mu _{s}& =\frac{1}{3}\mu _{B}-\frac{1}{3}\mu _{Q}-\mu _{S}
\end{align}

Using Eq.\thinspace (\ref{hadron2quark}) we can rewrite the observable
susceptibilities in terms of quark susceptibilities, which are derivatives
of the NJL potential. For example, the baryon susceptibility is 
\end{subequations}
\begin{align}
\chi _{B}=& \frac{\partial ^{2}\ln \mathcal{Z}}{\partial \mu _{B}^{2}}%
=\left( \frac{1}{3}\frac{\partial }{\partial \mu _{q}}+\frac{1}{3}\frac{%
\partial }{\partial \mu _{s}}\right) ^{2}\ln \mathcal{Z}  \notag
\label{chiBchiquark} \\
=& \frac{1}{9}\chi _{qq}+\frac{2}{9}\chi _{qs}+\frac{1}{9}\chi _{ss}
\end{align}%
%
%
%
%
Since net strangeness is conserved and equal to zero $\langle N_{s}\rangle
=0 $ independent of the $u,d$ chemical potentials, 
\begin{equation}
\chi _{qs}=\frac{\partial }{\partial {\mu _{q}}}\langle N_{s}\rangle =0
\end{equation}%
so that the second term in Eq.\thinspace (\ref{chiBchiquark}) vanishes. 
Finally, we have 
\begin{equation}
\chi _{B}=\frac{1}{9}\chi _{qq}+\frac{1}{9}\chi _{ss}
\end{equation}%
Similarly, the charge susceptibility simplifies 
\begin{equation}
\chi _{Q}=\frac{1}{36}\chi _{qq}+\chi _{II}+\frac{1}{9}\chi _{ss}
\end{equation}%
because the cross term $\propto \chi _{us}$ is zero. Note the presence of
the strange susceptibility. This shows that both the charge and baryon
fluctuations are dominated by the light quark fluctuations, as we expect 
\cite{Hatta:2003wn}. The strange-hadron susceptibility is 
\begin{equation}
\chi _{S}=\chi _{s}
\end{equation}%
%
%
%
%

\section{Susceptibilities with Flavor}\label{Sec-qIs}

In Ref. \cite{Chen:2014ufa}, the explicit expressions for the quark (baryon)
susceptibilities $\chi _{B}^{(3)},\chi _{B}^{(4)}$\ in terms of
autocorrelation functions of the $\sigma $\ zero mode $\langle \sigma
^{n}\rangle $\ were derived for the GN model. The $k$-th order susceptibility
requires only correlation functions of $k$-th order and lower, see Eqs. (6),\
(7),\ and (8) in \cite{Chen:2014ufa}. While higher order susceptibilities
typically diverge with larger powers of the correlation length $\xi $ \cite%
{Stephanov:2008qz}, we found that the corresponding autocorrelation function 
$\langle \sigma ^{n}\rangle $ does \emph{not} always provide the complete
leading order contribution to $\chi _{n}$. It also implies that the
evaluation of $\chi _{n}$ can be organized diagrammatically corresponding to
a finite set of terms in perturbation theory. As we extend the analysis to
NJL, the diagrammatic expansion becomes more complicated, however, we can
use it to help explain the behaviour of strange quark susceptibilities near
the critical point.

When writing these susceptibilities in terms of order parameter
fluctuations, we must include three condensates, one for each flavor, and
hence three fluctuation fields $\sigma_\alpha$, $\alpha=u,d,s$. Due to the
anomaly-induced six-quark interaction vertex, the $\sigma$ ``correlator''
(statistical covariance) $\partial^2\Omega/\partial\sigma_\alpha\partial%
\sigma_\beta$ is not diagonal in flavor. This fact is derived and a brief
description of the diagrammatic system is found in Appendix \ref%
{app:diagrams}.

We have checked numerically that the expansion in terms of $\sigma$
correlators is equivalent to taking the $\mu$ derivatives directly at the
minimum of the effective potential by brute force, e.g. a finite difference
method. The $\sigma$-correlator diagrams provide an easier method to
organize and analyze the many terms arising from the evaluation of higher
order and multi-flavor susceptibilities.

\begin{figure}[h]
\includegraphics[width=0.5\textwidth]{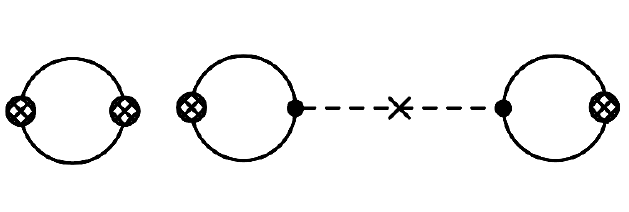}
\caption{Diagrams (a) and (b) respectively correspond to the first and second terms in
Eq.\thinspace (\protect\ref{chi2}). Each crossed circle
denotes an insertion of the $\bar{q}_{\protect\alpha }q_{\protect\alpha }$\
operator arising from taking a $\protect\mu _{\protect\alpha }$\ derivative
to the pressure which is proportional to the logarithm of the partition
function. The solid and dashed lines denote the quark and the $\protect%
\sigma _{\protect\alpha }$\ propagators. The cross denotes the flavor mixing
of the $\protect\sigma $\ propagators.}
\label{fig:suscept2}
\end{figure}

Using the diagrammatic system developed in Appendix \ref{app:diagrams}, we
can quickly identify large contributions to flavor-dependent observables. As
mentioned above, the critical mode is associated with light quarks, since
this transition occurs at smaller $\mu_B$. Therefore the most divergent
contributions are $n$-point correlation functions of the $\sigma_q$
fluctuations.

\begin{figure}[tbp]
\includegraphics[width=0.5\textwidth]{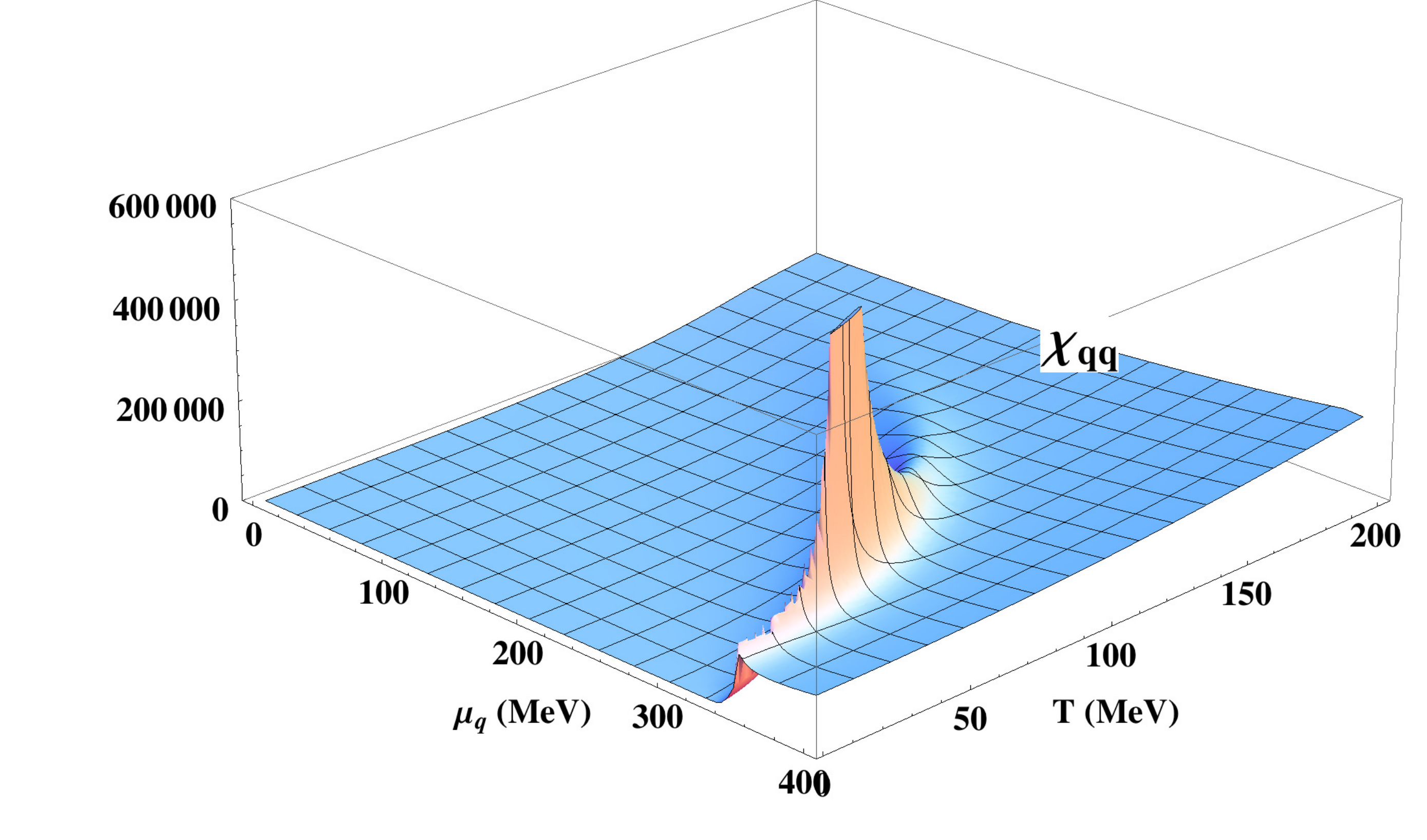} %
\includegraphics[width=0.5\textwidth]{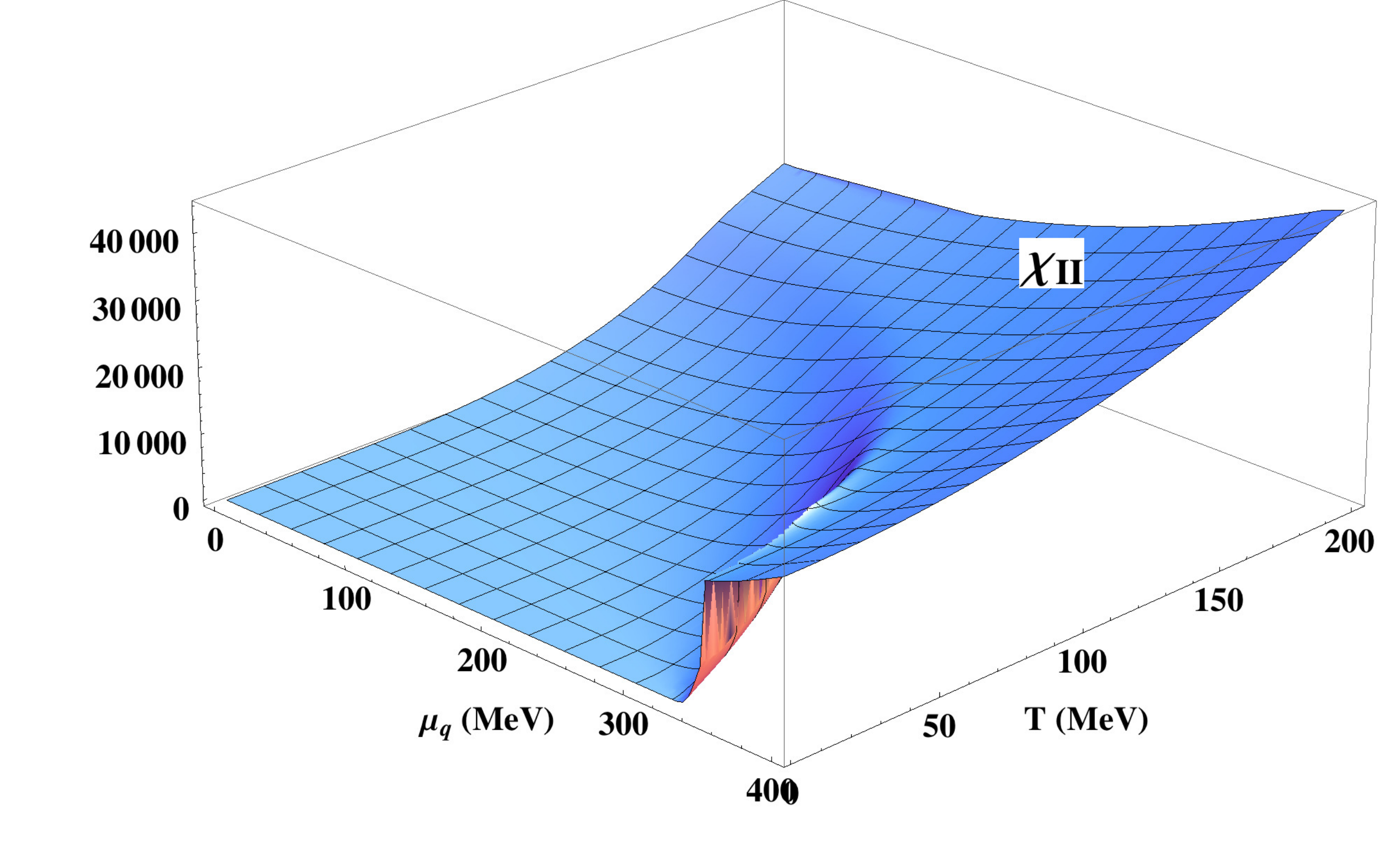} %
\includegraphics[width=0.5\textwidth]{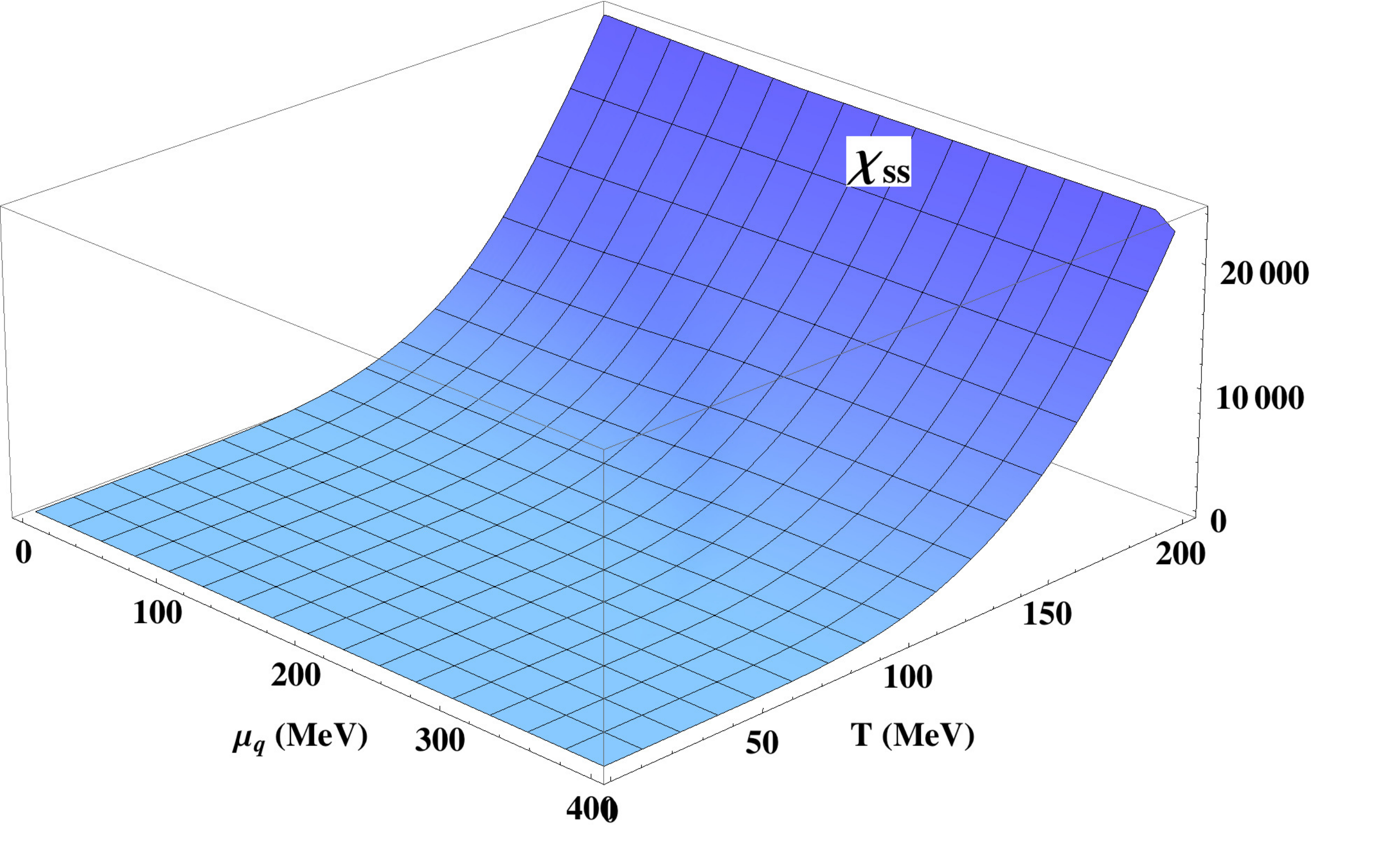}
\caption{$\protect\chi_{qq},\protect\chi_{II}, \protect\chi_{ss}$, three
independent components of the second order susceptibilities for the
conserved charge $(q,I,s)$. }
\label{fig:chi2-qIs}
\end{figure}

The diagrams shown in Figure \ref{fig:suscept2} explain the behaviour of the
second order susceptibilities $\chi _{II}$ and $\chi _{ss}$. Since the first
derivatives with respect to $\mu _{I}$ and $\mu _{s}$ vanish identically at $%
\mu _{I},\mu _{s}=0$, the diagram on the right vanishes because it has only
a single $\mu $-derivative acting on the distribution function. As a result,
there are no contributions involving $\sigma $-correlators to these
susceptibilities, and they show no divergence.

\begin{figure}[tbp]
\includegraphics[width=0.5\textwidth]{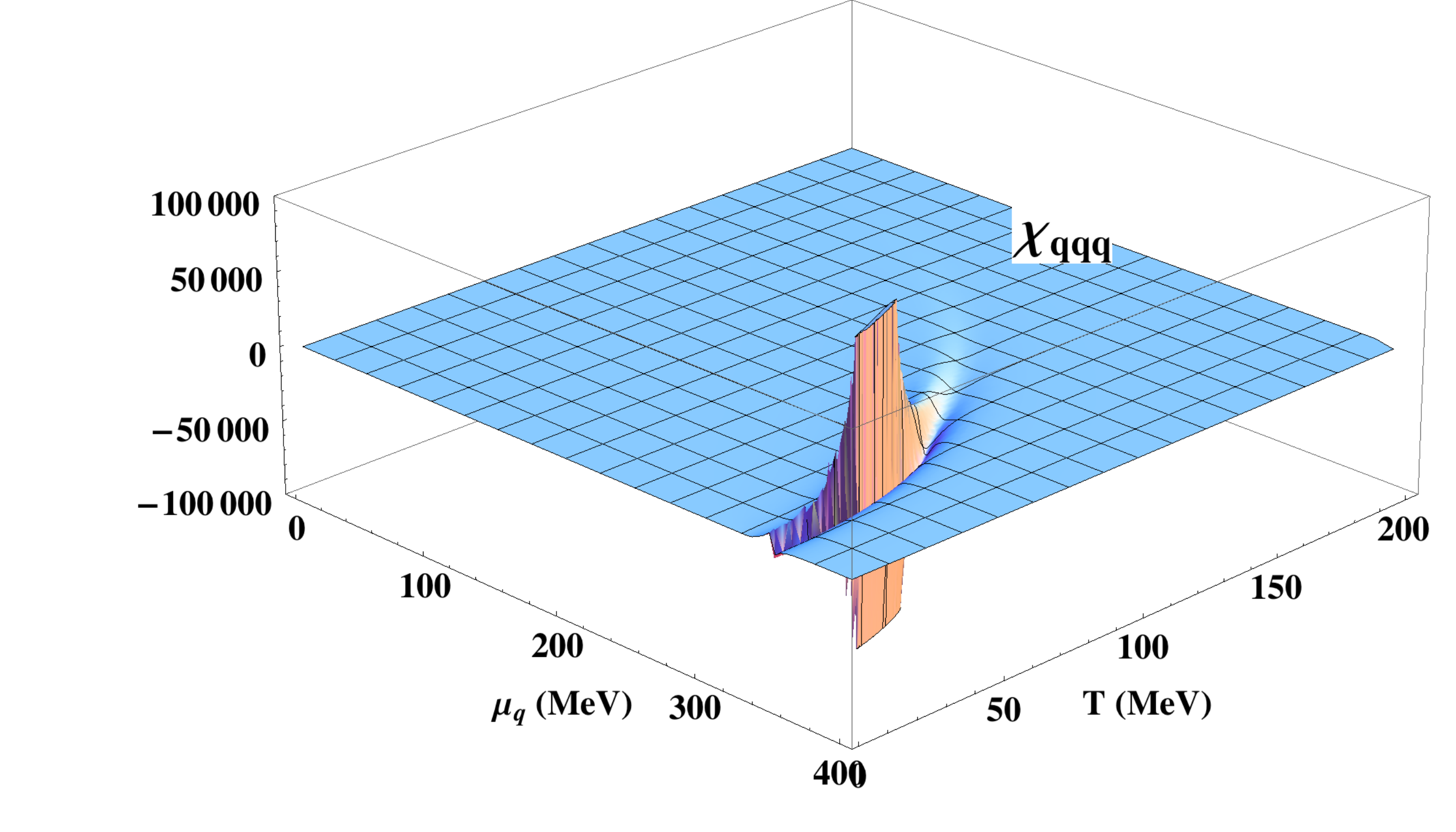}\newline
\includegraphics[width=0.5\textwidth]{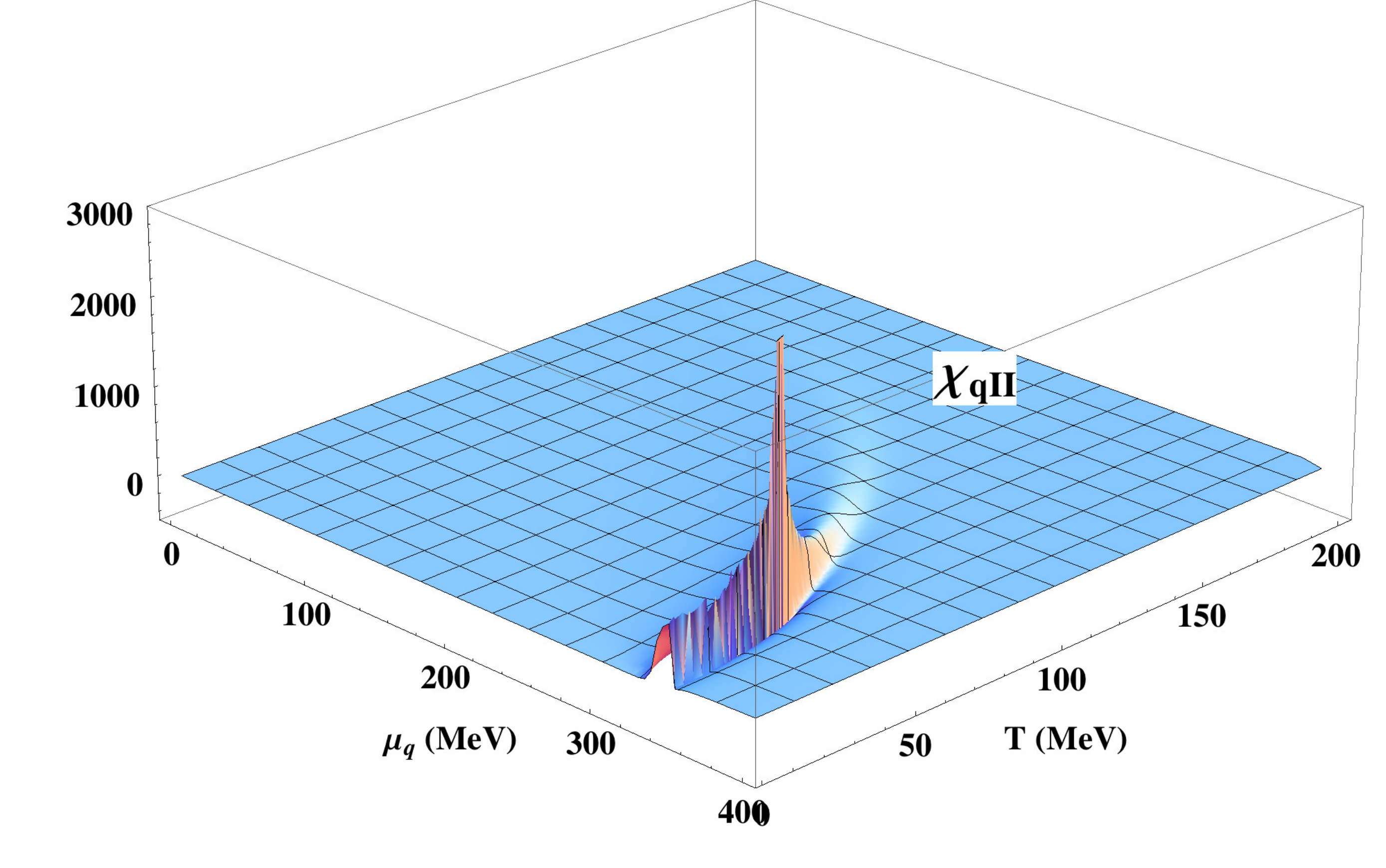}\newline
\includegraphics[width=0.5\textwidth]{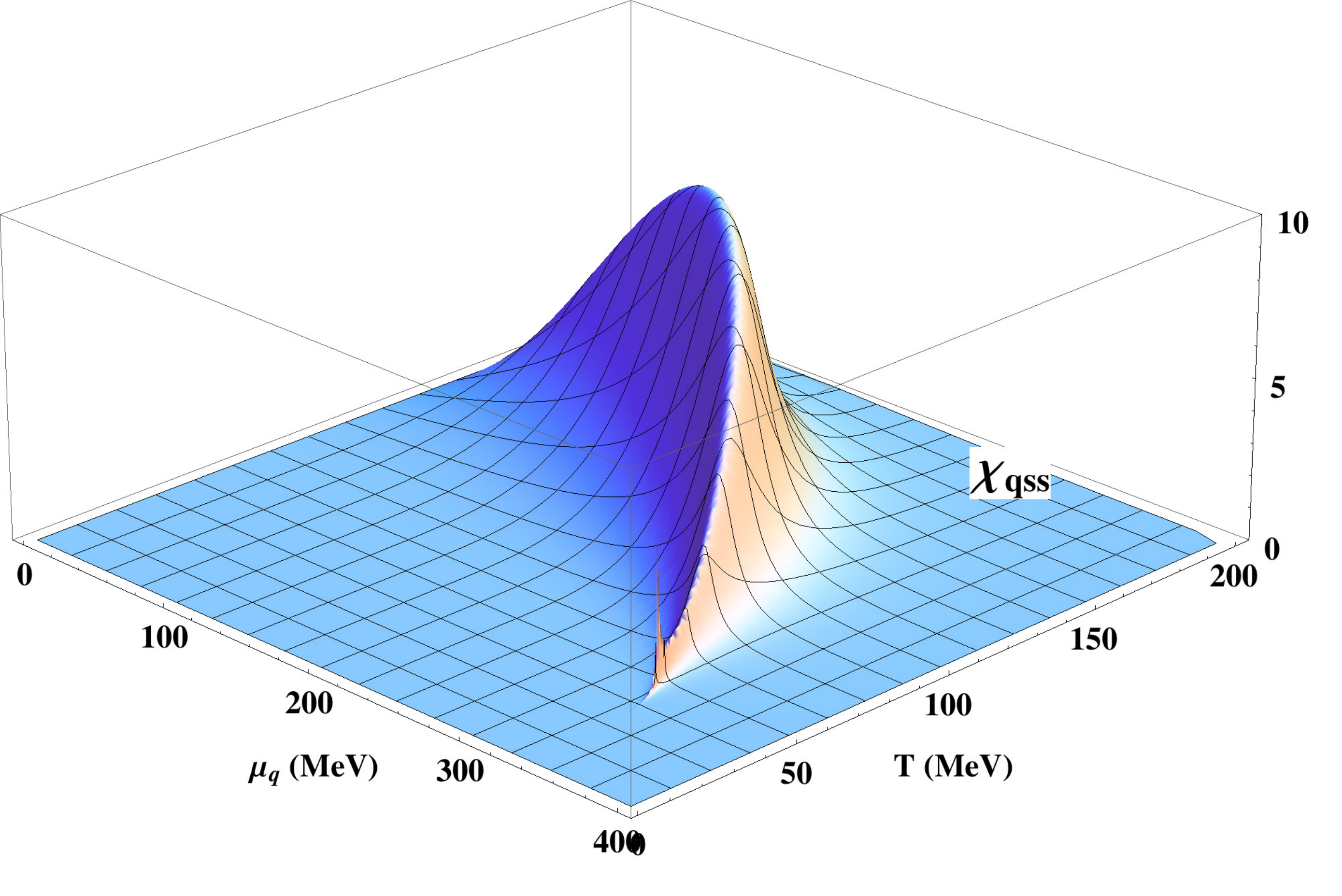}
\caption{T$\protect\chi_{qqq},\protect\chi_{qII}, \protect\chi_{qss}$, three
independent components of the third order susceptibilities for the conserved
charge $(q,I,s)$. }
\label{fig:chi3-qIs}
\end{figure}

\begin{figure}[tbp]
\includegraphics[width=0.5\textwidth]{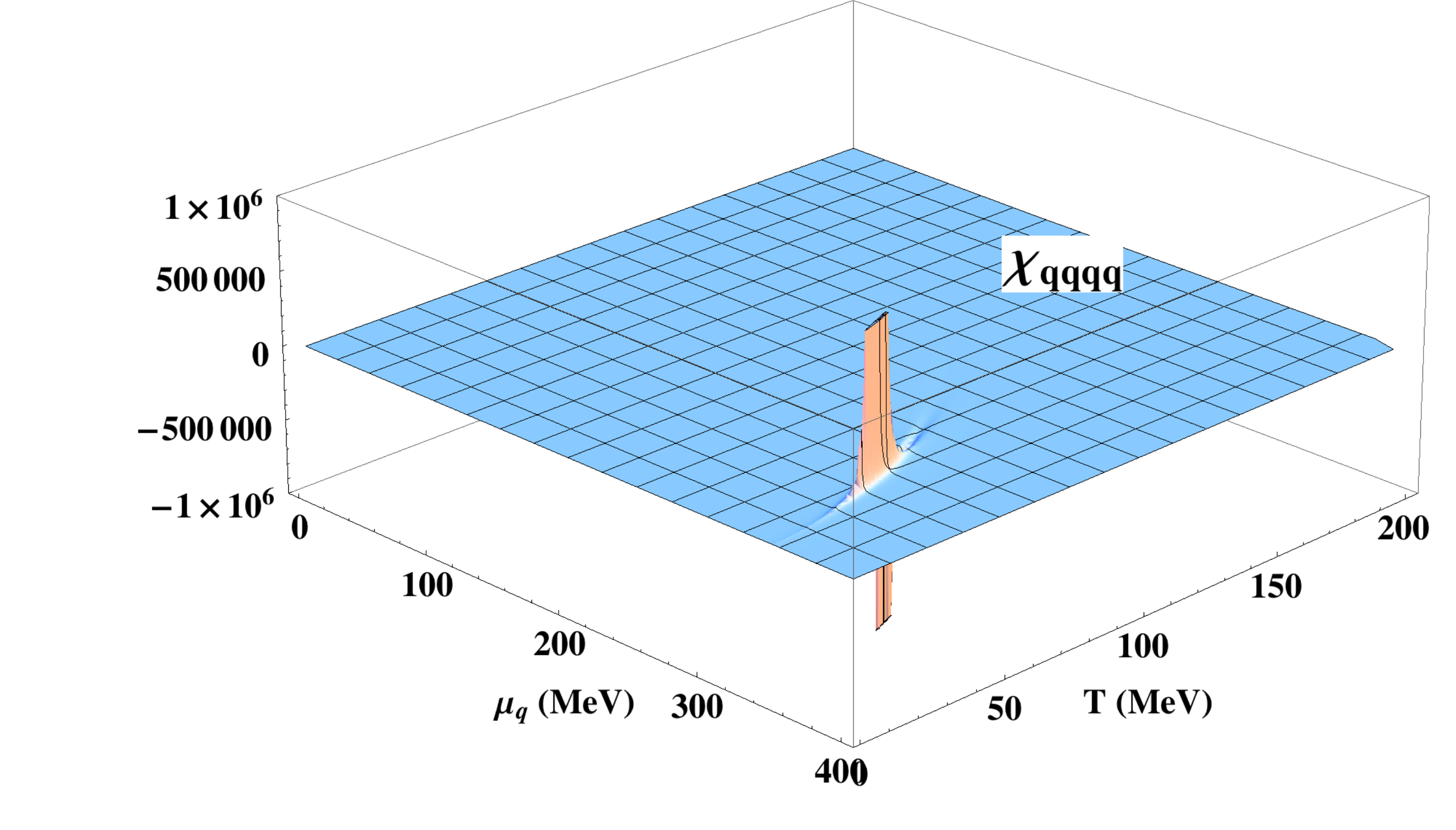}\newline
\includegraphics[width=0.5\textwidth]{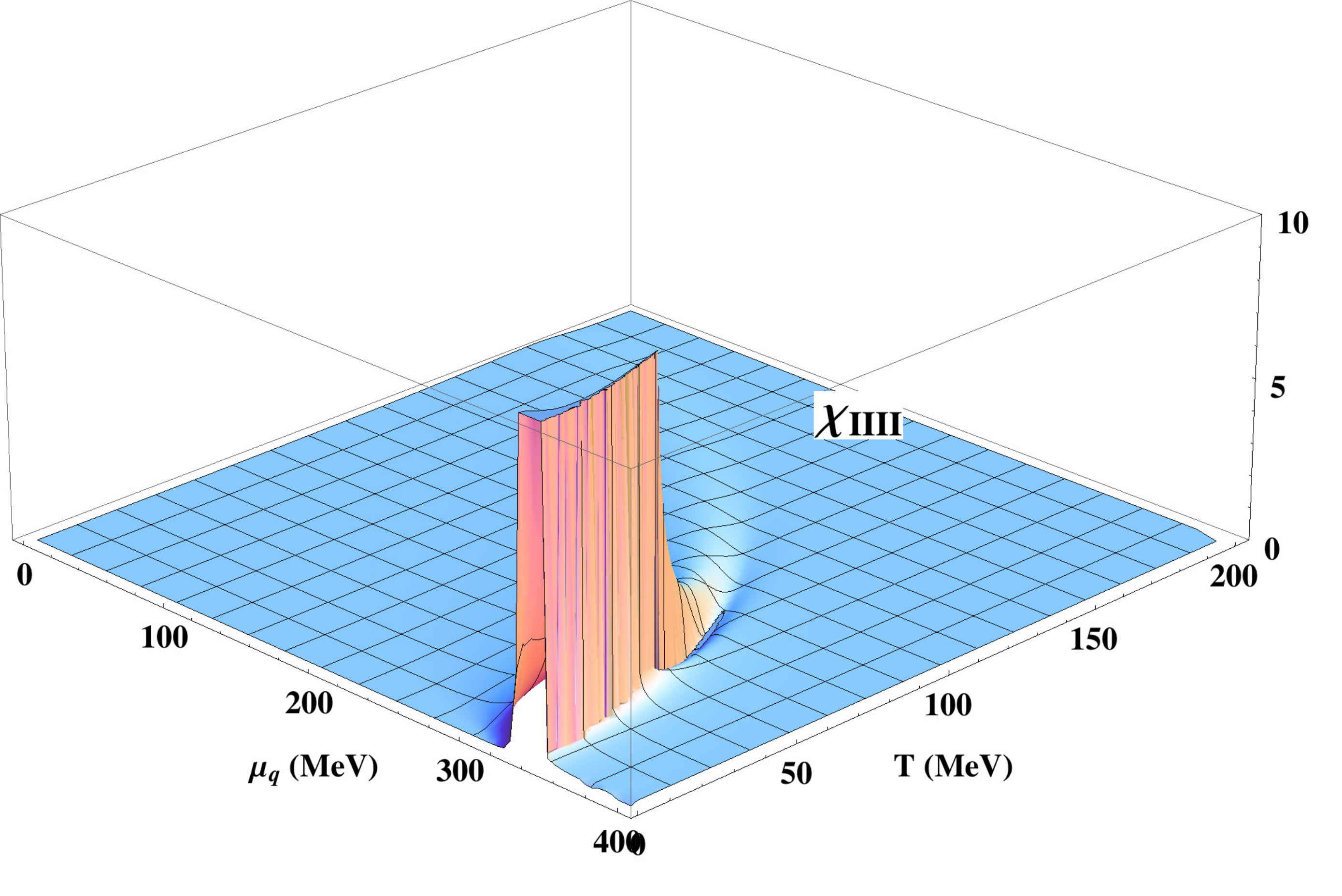}\newline
\includegraphics[width=0.5\textwidth]{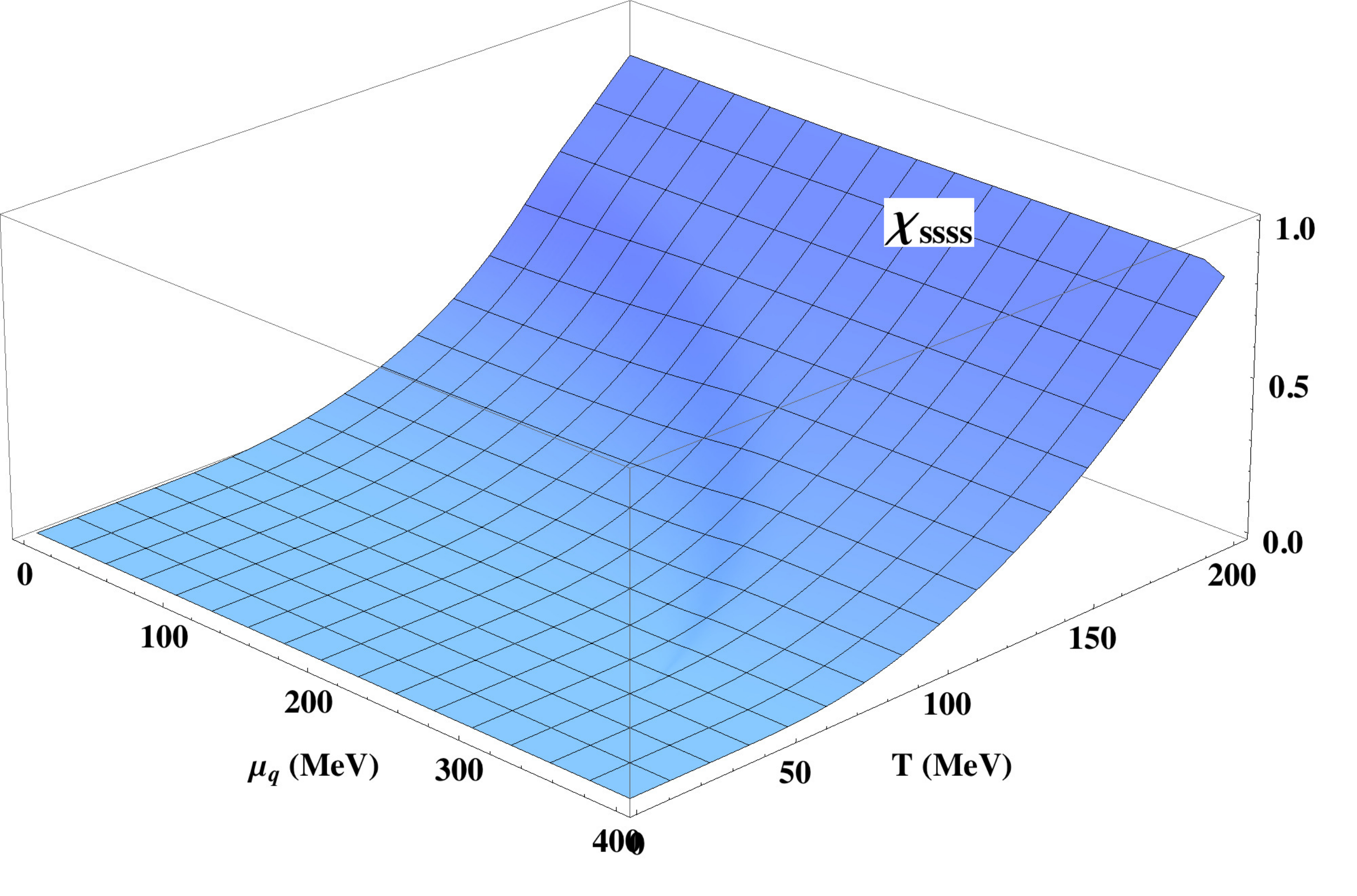}
\caption{The fourth-order susceptibilities, $\protect\chi_{q}^{(4)},\protect\chi%
_{I}^{(4)}$ and $\protect\chi_{s}^{(4)}$. }
\label{fig:chi4-qIs-1}
\end{figure}

\begin{figure}[tbp]
\includegraphics[width=0.5\textwidth]{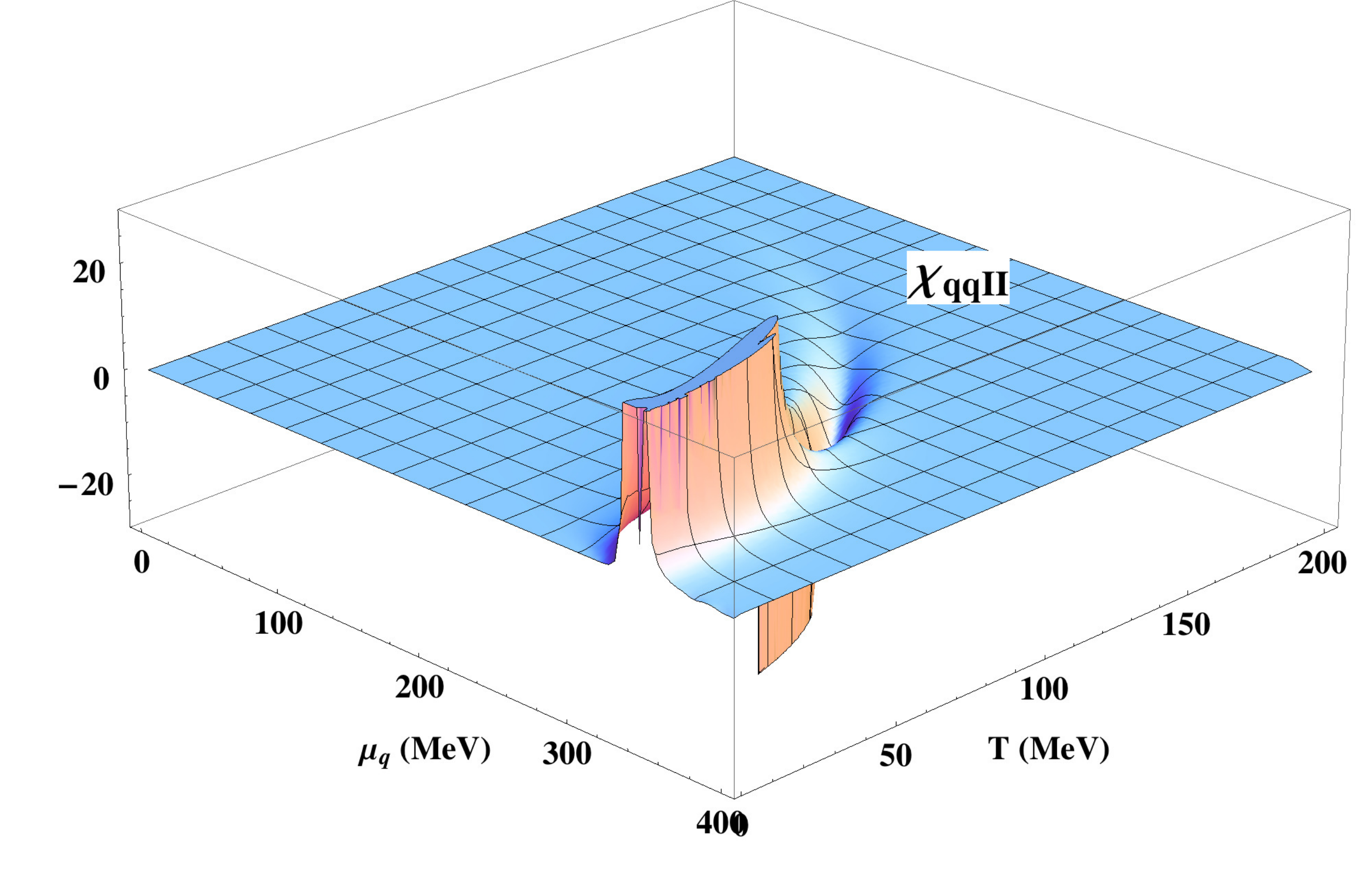}\newline
\includegraphics[width=0.5\textwidth]{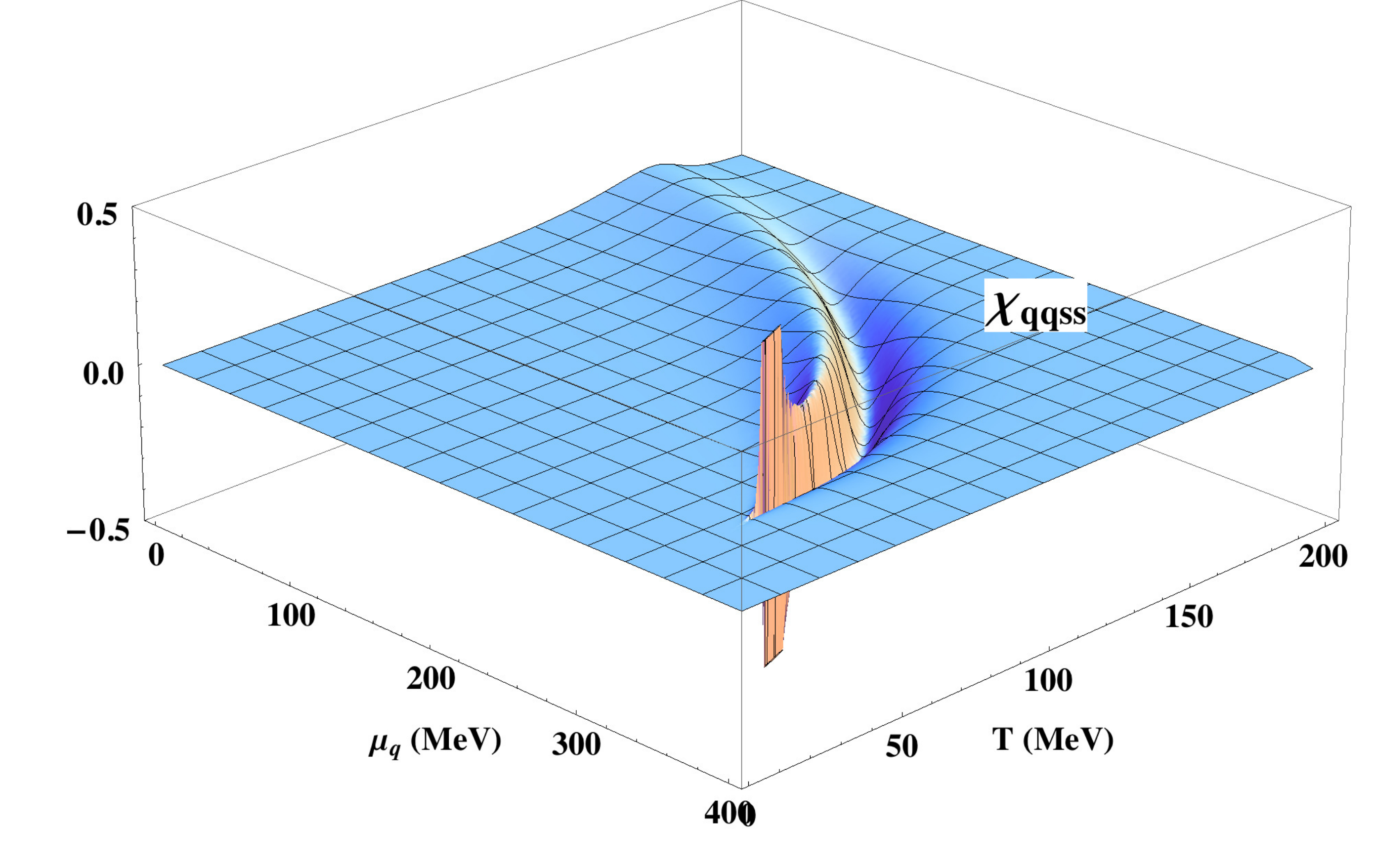}\newline
\includegraphics[width=0.5\textwidth]{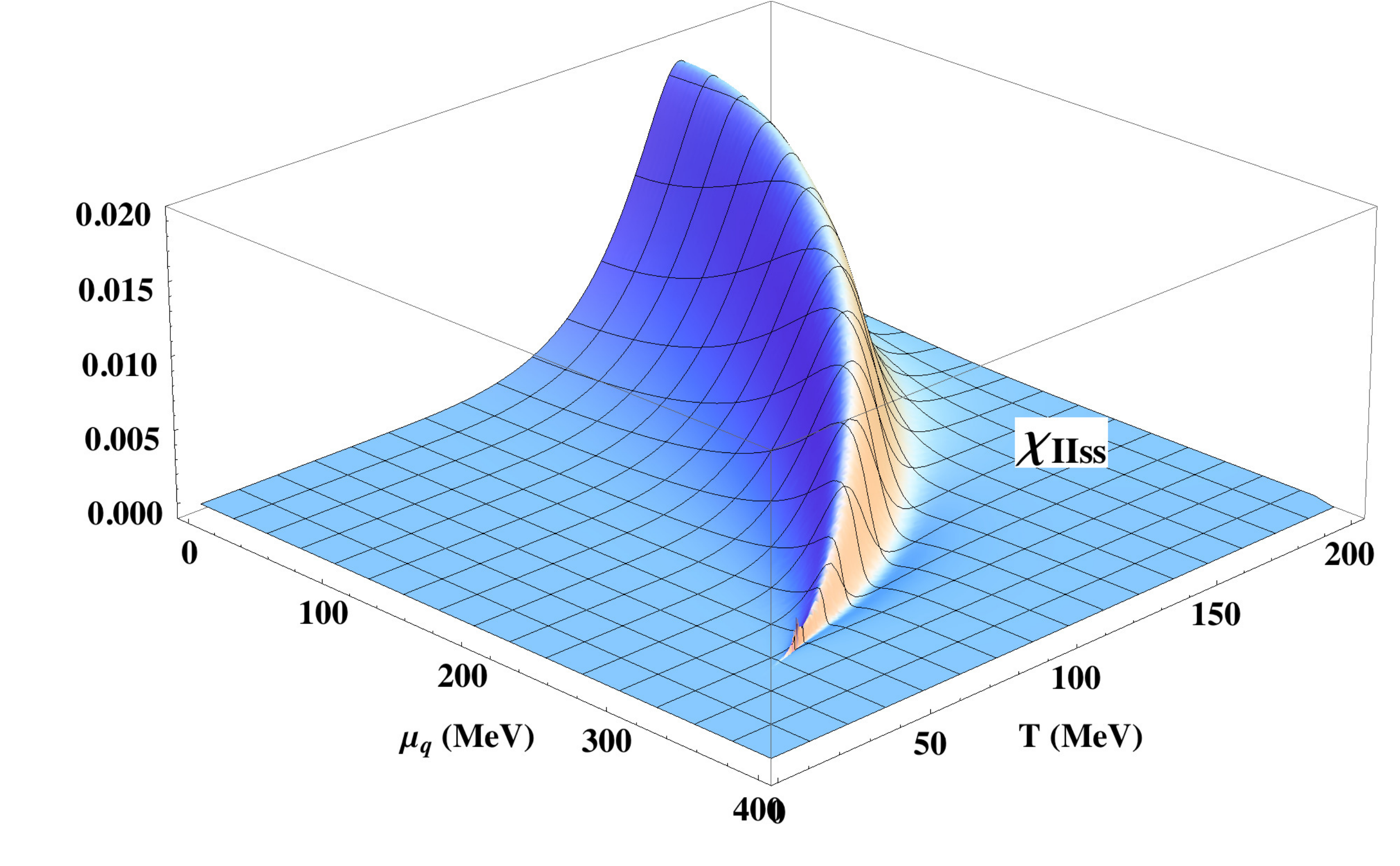}\newline
\includegraphics[width=0.5\textwidth]{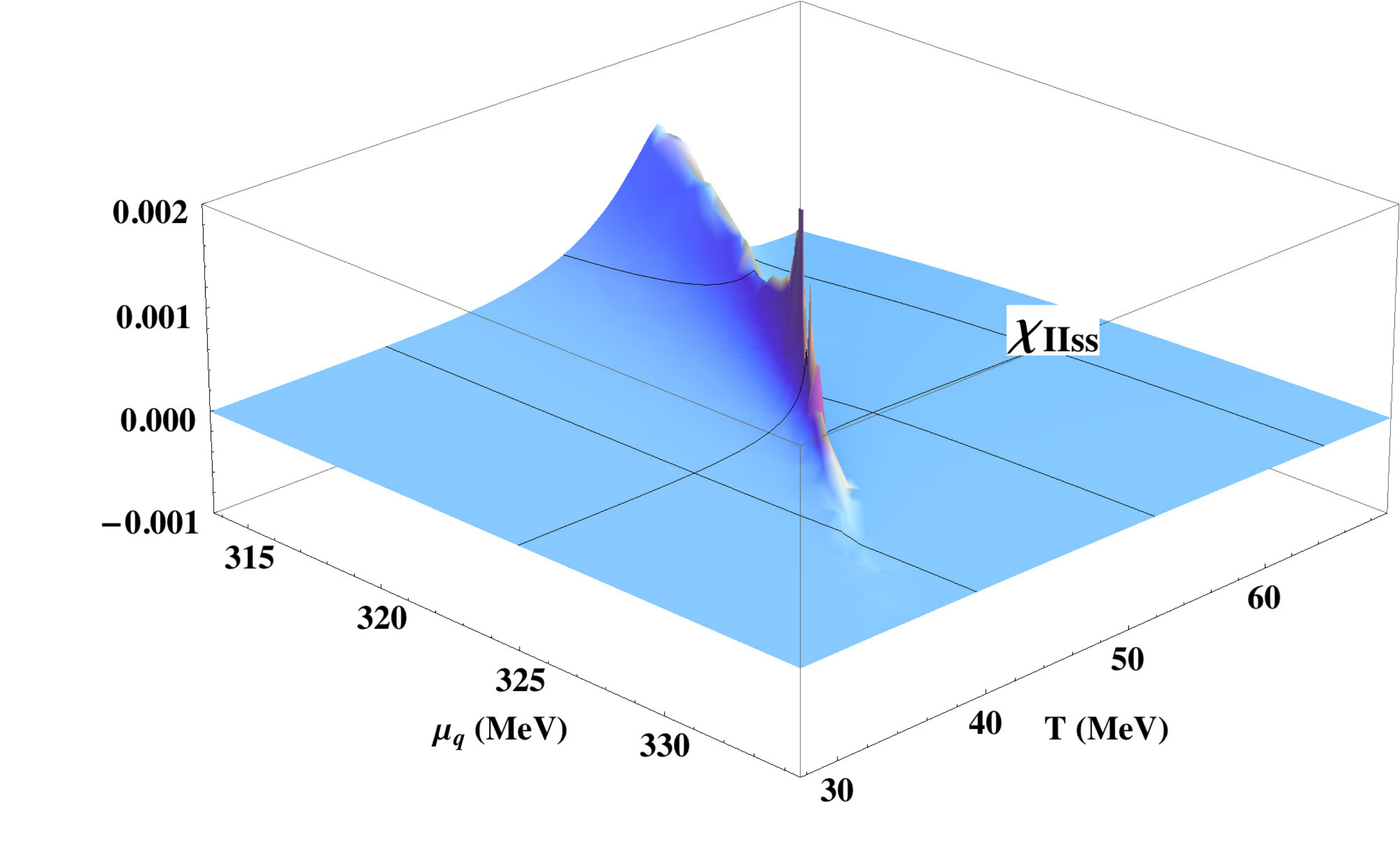}
\caption{The fourth-order susceptibilities, $\protect\chi_{qqII},\protect\chi%
_{qqss}$ and $\protect\chi_{IIss}$. }
\label{fig:chi4-qIs-2}
\end{figure}

In Figures \ref{fig:chi2-qIs}, \ref{fig:chi3-qIs}, \ref{fig:chi4-qIs-1} and %
\ref{fig:chi4-qIs-2} we display the complete set of quark susceptibilities $%
\chi_{\alpha\beta...}$ up to the fourth derivatives, from which we need to
construct the fluctuations of observables.

\begin{figure}[tbp]
\includegraphics[width=150pt]{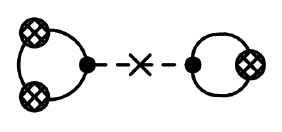}
\caption{Diagram contributing the singularity to $\protect\chi_{qII}$ and $%
\protect\chi_{qss}$. }
\label{fig:chiqss}
\end{figure}

The third order susceptibilities are straightforwardly understood since each
is a single $\mu_q$ derivative on a second order susceptibilities. Thus, we
can confirm by eye the correct behaviours: $\chi_{qqq}$ is odd across the
first order line, corresponding to the peak seen in $\chi_{qq}$. $\chi_{qII}$
has a peak around the first-order line, corresponding to the jump in $%
\chi_{II}$ but remains positive at high $\mu$ since $\chi_{II}$ continues
with positive derivative. The variation of $\chi_{ss}$, exhibited by nonzero 
$\chi_{qss}$, is too small in magnitude to see in Fig. \ref{fig:chi2-qIs}.
The derivative with $\mu_q$ contributes a singularity at the CEP to each $%
\chi_{qII}$ and $\chi_{qss}$, which arises from a new nonzero diagram, shown
in Fig. \ref{fig:chiqss}.

Among the fourth order susceptibilities, three $\chi_{q}^{(4)},\chi_{qqII},%
\chi_{qqss}$ can be understood in the same way as the third order
susceptibilities looking at the $\mu_q$-derivatives of the previous plots.
For the other three, we can use the diagrammatic analysis.

\begin{figure}[tbp]
\includegraphics[width=0.5\textwidth]{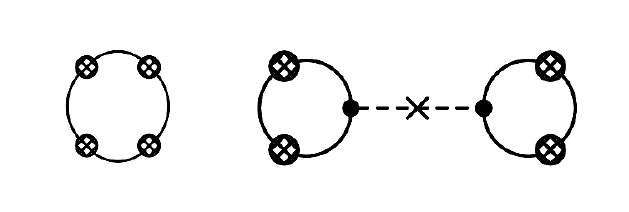}
\caption{Diagrams in the expansion of $\protect\chi_{s}^{(4)}$. Due to the $%
\protect\mu_s\leftrightarrow-\protect\mu_s$ symmetry, the only nonvanishing
diagram involving a $\protect\sigma$ field correlator is at right, and can
only involve the $\protect\sigma_s$ order parameter field. }
\label{fig:chissss}
\end{figure}

$\chi _{s}^{(4)}$ can have a singular contribution. Again, the diagrams
contributing are severely constrained by the fact that odd derivatives in $%
\mu _{s}$ vanish. The only two diagrams with an even of number of $\mu _{s}$
derivatives on the external legs are shown in Figure \ref{fig:chissss}. The
diagram on the right does contain a $\sigma $ correlator, which is
flavor-diagonal, connecting one $s$ distribution function (loop) to a
second. This correlator is also singular at the light quark critical point,
due to the flavor coupling term in the NJL Lagrangian. This fact is
recognized in many studies of NJL \cite{Fu:2010ay}, that the strange
condensate also has a small discontinuity across the first order line.
However, the singularity at the CEP is suppressed quantitatively by the
small discontinuity and large strange quark bare mass. With high precision
numerics, it becomes visible in Figure \ref{fig:m2} below.

The same diagrams in Fig. \ref{fig:chissss} contribute to $\chi_{I}^{(4)}$.
With the presence of $\sigma $-correlators for the light quarks, we can
expect some singularity, larger than in $\chi _{s}^{(4)}$ but smaller than in $%
\chi _{q}^{(4)}$ due to some cancellation of leading terms. For $\chi _{IIss}$
only the diagram on the right in Fig. \ref{fig:chissss} contributes (because
the $\left( \bar{q}_{I}q_{I}\right) ^{2}\left( \bar{q}_{s}q_{s}\right) ^{2}$%
\ operators cannot be contracted to a single loop).  The $\sigma$-correlator implies $\chi_{IIss}$ is singular, but again the divergence is relevant only in a very small region, even accounting for the $\xi$-scaling of the coefficients weighting each $\sigma$-correlator contribution.

\begin{figure}[tbp]
\includegraphics[width=0.5\textwidth]{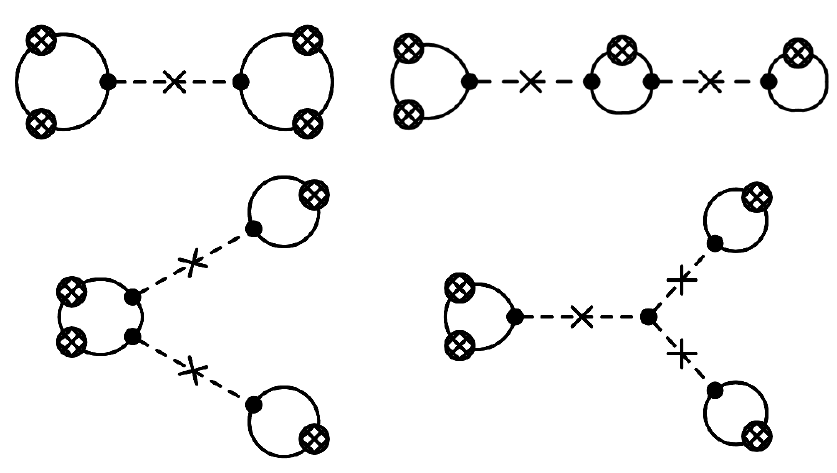}
\caption{Diagrams in the expansion of $\protect\chi_{uuss}$ }
\label{fig:chi-uuss}
\end{figure}

The cross-correlation in flavor $\chi _{qqss}$ receives a larger singular contribution. To see this,
consider the four possible diagrams shown in Figure \ref{fig:chi-uuss}. In
each diagram, there are at least two $\sigma _{q}$ propagators, which
together give a factor $\propto \xi ^{4}$. Further, we can estimate that the
dominant effect near the CEP is the diagram with the $\sigma _{u}$ 3-point
function: involving 3 $\sigma $ correlators (each scaling as $\xi ^{2}$) and
the 3-point vertex (scaling as $(\xi T)^{-1.5}$, obtained by solving the gap
equation near the CP), the diagrams has an overall scaling $\sim \xi ^{4.5}$%
, before the $\xi $-dependence of the coefficient is applied. The dominance
of this term is confirmed numerically. However, the overall magnitude of the
peak is exponentially suppressed by the strange mass as compared to the
singularities in light quark susceptibilities, which we have checked
by varying the bare mass of strange quark.

Writing out the nongaussian hadron-level susceptibilities, as expected, only
3 third-order susceptibilities and 6 fourth-order susceptibilities are
linearly independent. The third order susceptibilities are 
\begin{subequations}
\label{3susceptibilities}
\begin{align}
\chi _{B}^{(3)}& =\frac{1}{27}\chi _{q}^{(3)}+\frac{1}{9}\chi _{qss} \\
\chi _{Q}^{(3)}& =\frac{1}{216}\chi _{q}^{(3)}+\frac{1}{2}\chi _{qII}+\frac{1%
}{18}\chi _{qss} \\
\chi _{S}^{(3)}& =0 \\
\chi _{BBQ}& =\frac{1}{54}\chi _{q}^{(3)}-\frac{1}{18}\chi _{qss} \\
\chi _{BBS}& =-\frac{2}{9}\chi _{qss} \\
\chi _{BQQ}& =\frac{1}{108}\chi _{qqq}+\frac{1}{3}\chi _{qII} \\
\frac{1}{6}\chi _{BSS}&=\chi _{BQS} =\frac{1}{3}\chi _{QSS}=\frac{1}{2}\chi
_{QSS}=\frac{1}{18}\chi _{qss}
\end{align}%
\end{subequations}
and the fourth order susceptibilities are 
\begin{subequations}
\label{4susceptibilites}
\begin{align}
\chi _{B}^{(4)}& =\frac{1}{81}\chi _{q}^{(4)}+\frac{2}{27}\chi _{qqss}+\frac{%
1}{81}\chi _{s}^{(4)} \\
\chi _{Q}^{(4)}& =\frac{1}{1296}\chi _{q}^{(4)}+\frac{1}{54}\chi _{qqss}+%
\frac{1}{6}\chi _{qqII}  \notag \\
& ~~~+\chi _{I}^{(4)}+\frac{2}{3}\chi _{IIss}+\frac{1}{81}\chi _{s}^{(4)} \\
\chi _{S}^{(4)}& =\chi _{s}^{(4)} \\
\chi _{BBBQ}& =\frac{1}{162}\chi _{q}^{(4)}-\frac{1}{54}\chi _{qqss}-\frac{1%
}{81}\chi _{s}^{(4)} \\
\chi _{BBBS}& =-\frac{1}{9}\chi _{qqss}-\frac{1}{27}\chi _{s}^{(4)} \\
\chi _{BBQQ}& =\frac{1}{324}\chi _{q}^{(4)}-\frac{1}{108}\chi _{qqss}+\frac{1%
}{9}\chi _{qqII}  \notag \\
& ~~~+\frac{1}{9}\chi _{IIss}+\frac{1}{81}\chi _{s}^{(4)} \\
\chi _{BBSS}& =\frac{1}{9}\chi _{qqss}+\frac{1}{9}\chi _{s}^{(4)} \\
\chi _{BQQQ}& =\frac{1}{648}\chi _{q}^{(4)}+\frac{1}{108}\chi _{qqss}+\frac{1%
}{6}\chi _{qqII}  \notag \\
& ~~~-\frac{1}{3}\chi _{IIss}-\frac{1}{81}\chi _{s}^{(4)} \\
\chi _{BQQS}& =\frac{1}{36}\chi _{qqss}-\frac{1}{3}\chi _{IIss}-\frac{1}{27}%
\chi _{s}^{(4)} \\
\chi _{BQSS}& =\frac{1}{18}\chi _{qqss}-\frac{1}{9}\chi _{s}^{(4)} \\
\chi _{QQQS}& =\frac{1}{36}\chi _{qqss}+\chi _{IIss}+\frac{1}{27}\chi
_{s}^{(4)} \\
\chi _{QQSS}& =\frac{1}{36}\chi _{qqss}+\chi _{IIss}+\frac{1}{9}\chi
_{s}^{(4)} \\
\chi _{BBQS}& =-\frac{1}{9}\chi _{BSSS}=\frac{1}{9}\chi _{QSSS}=\frac{1}{27}%
\chi _{s}^{(4)}
\end{align}%
\end{subequations}
Based on the above analysis, the behaviour of these susceptibilities near a critical point (magnitude and shape) are dominated by the light quark susceptibilities, such as $\chi_q^{(3)}$ at third order and $\chi_q^{(4)}$ at fourth order. Only in observables without these two terms can the singularity due to the strange quark fluctuations be visible. We will see these effects in the following section.

\section{Characteristics of the Susceptibilities}

\begin{figure}[tbp]
\includegraphics[width=0.5\textwidth]{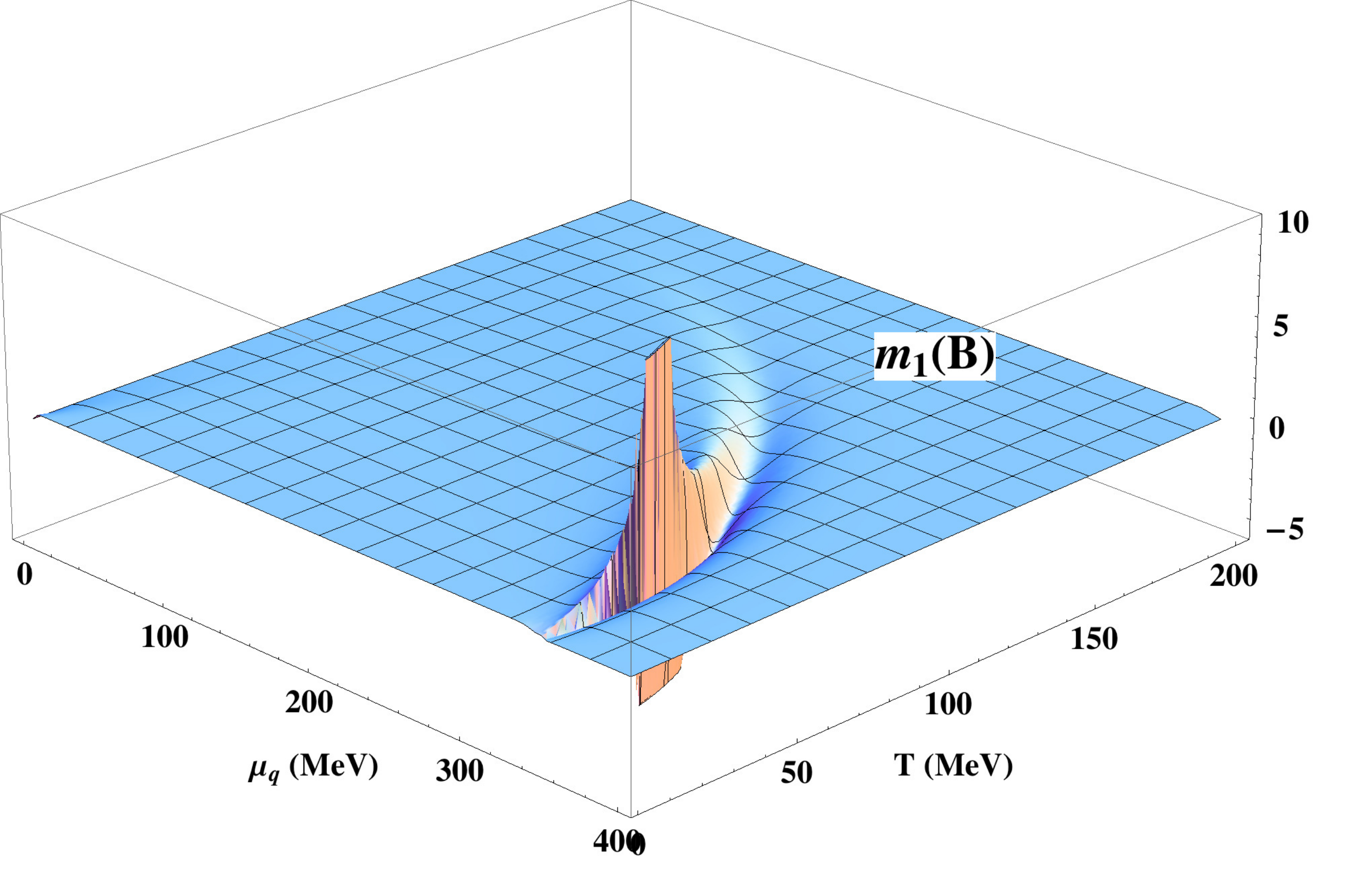}\newline
\includegraphics[width=0.5\textwidth]{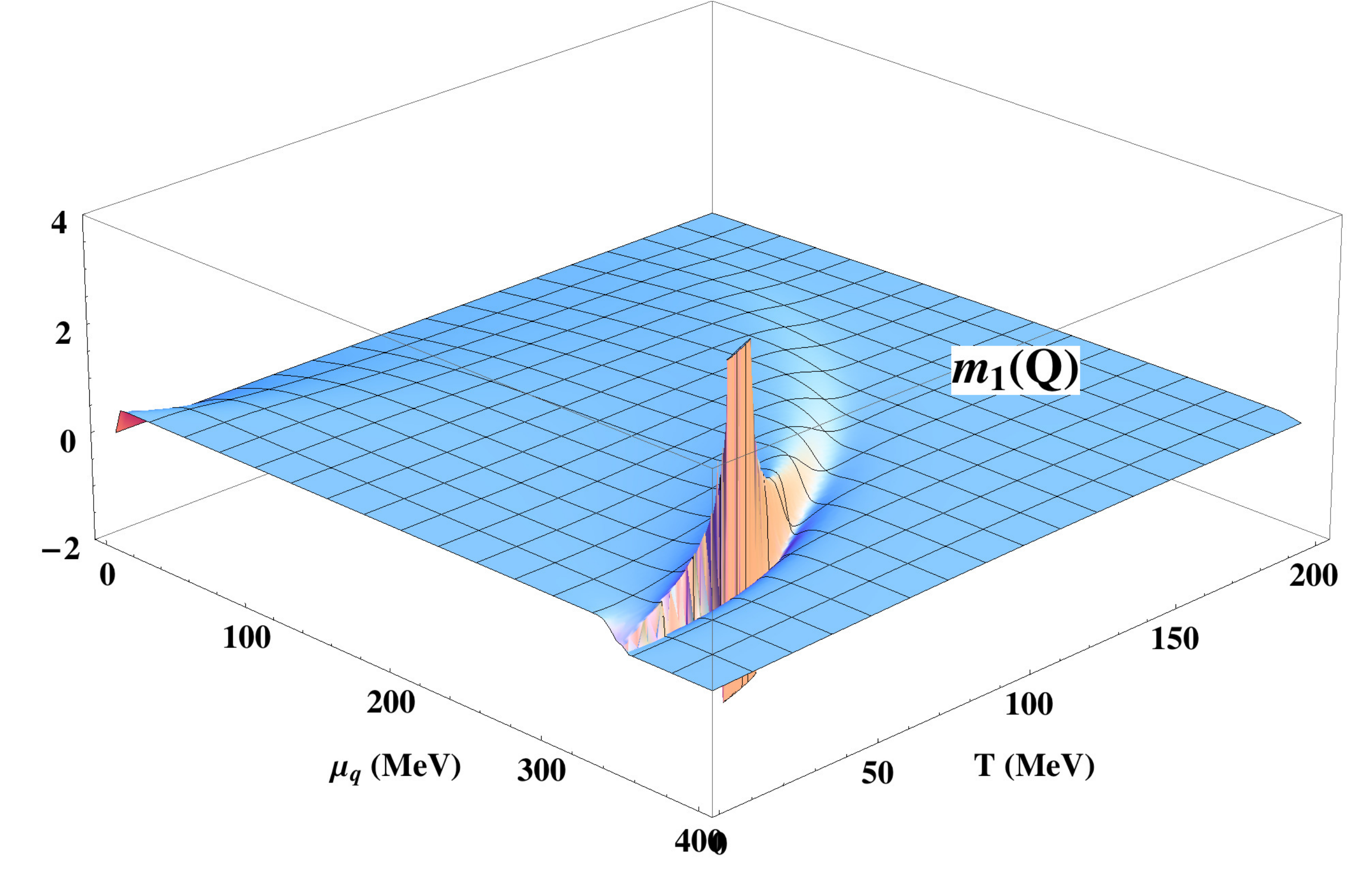}
\caption{$m_1$ for $B,Q$ on the phase diagram. $m_1$ for $S$ is zero. }
\label{fig:m1}
\end{figure}

\begin{figure}[tbp]
\includegraphics[width=0.5\textwidth]{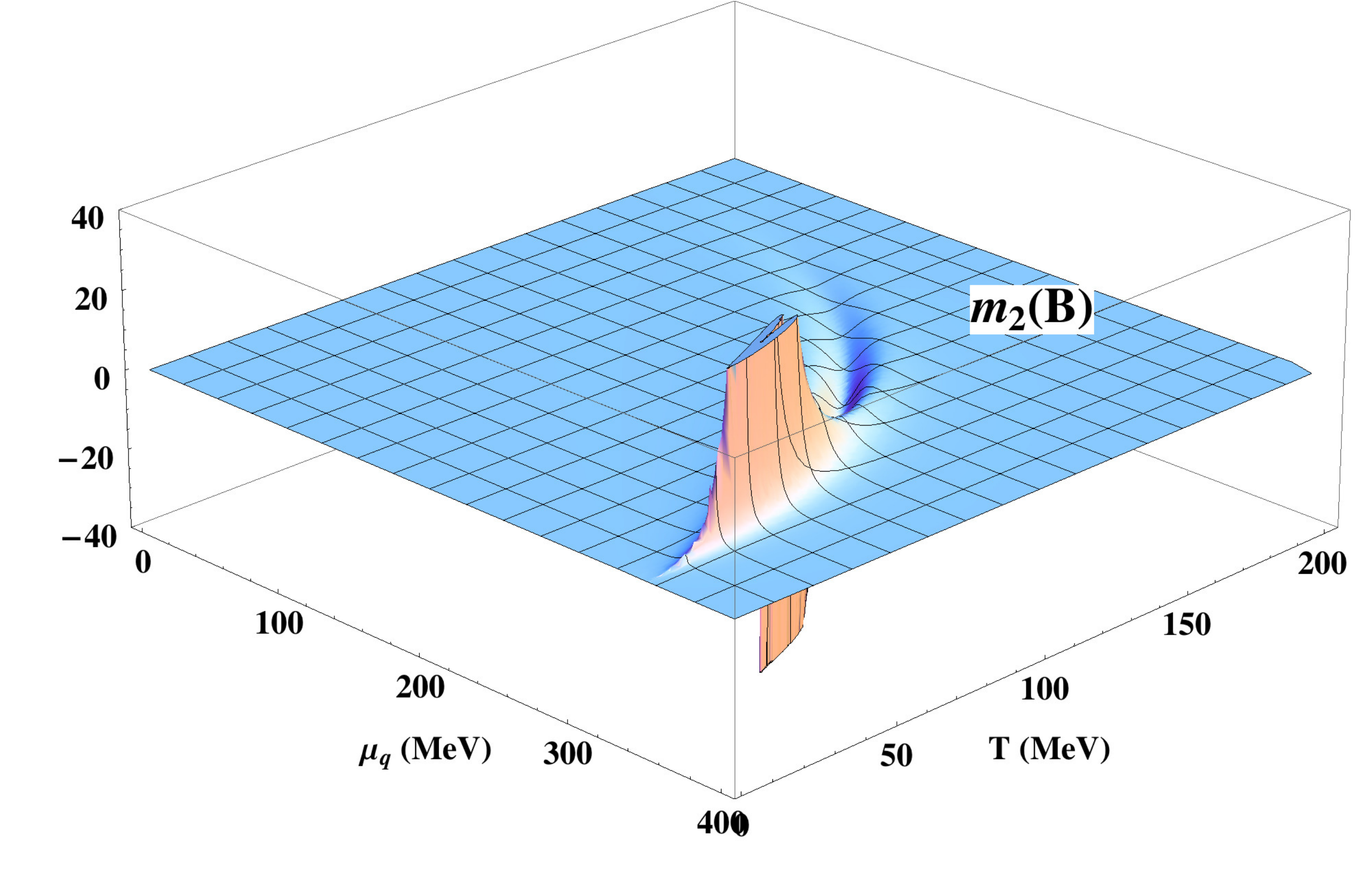}\newline
\includegraphics[width=0.5\textwidth]{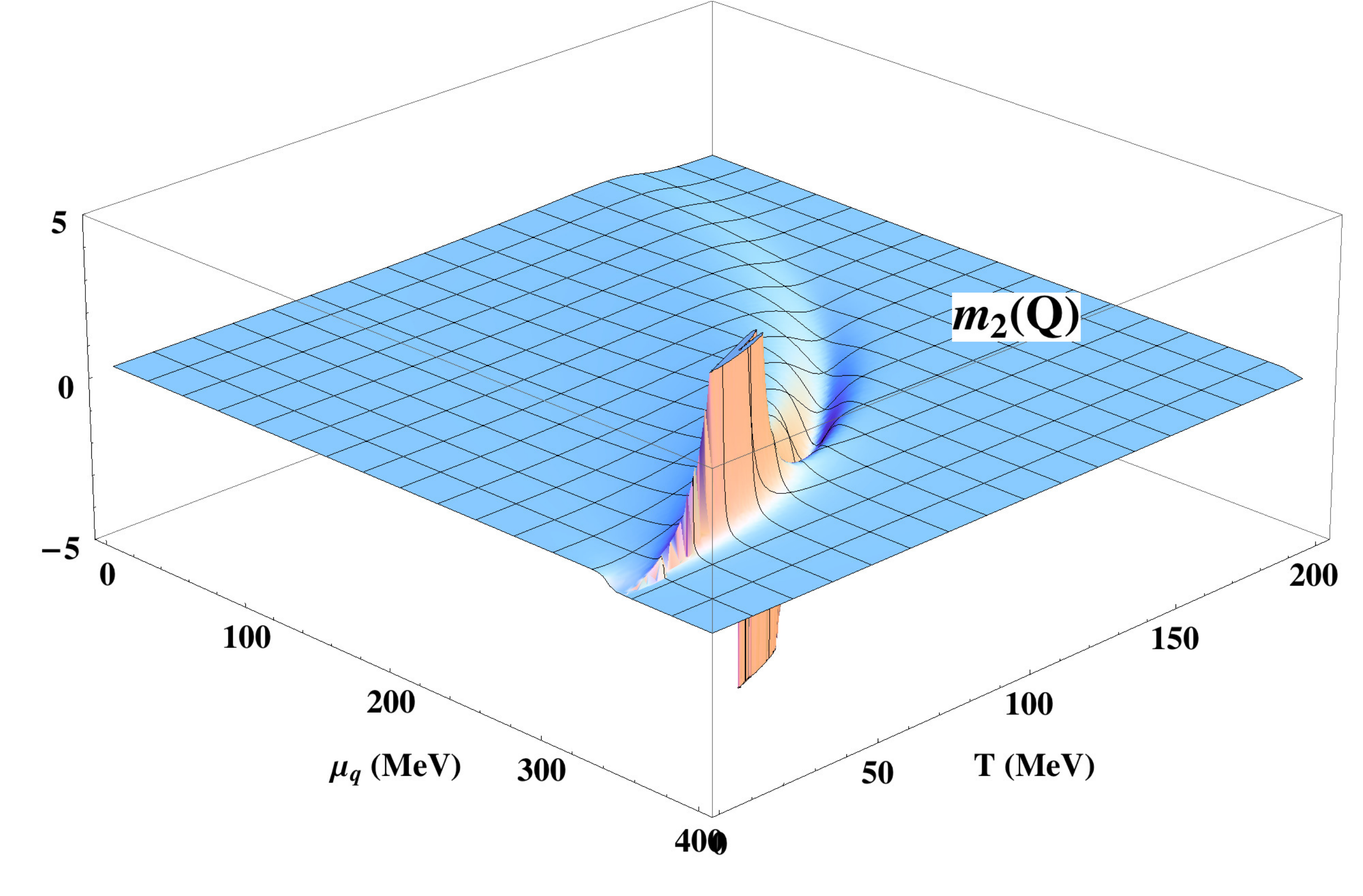}\newline
\includegraphics[width=0.5\textwidth]{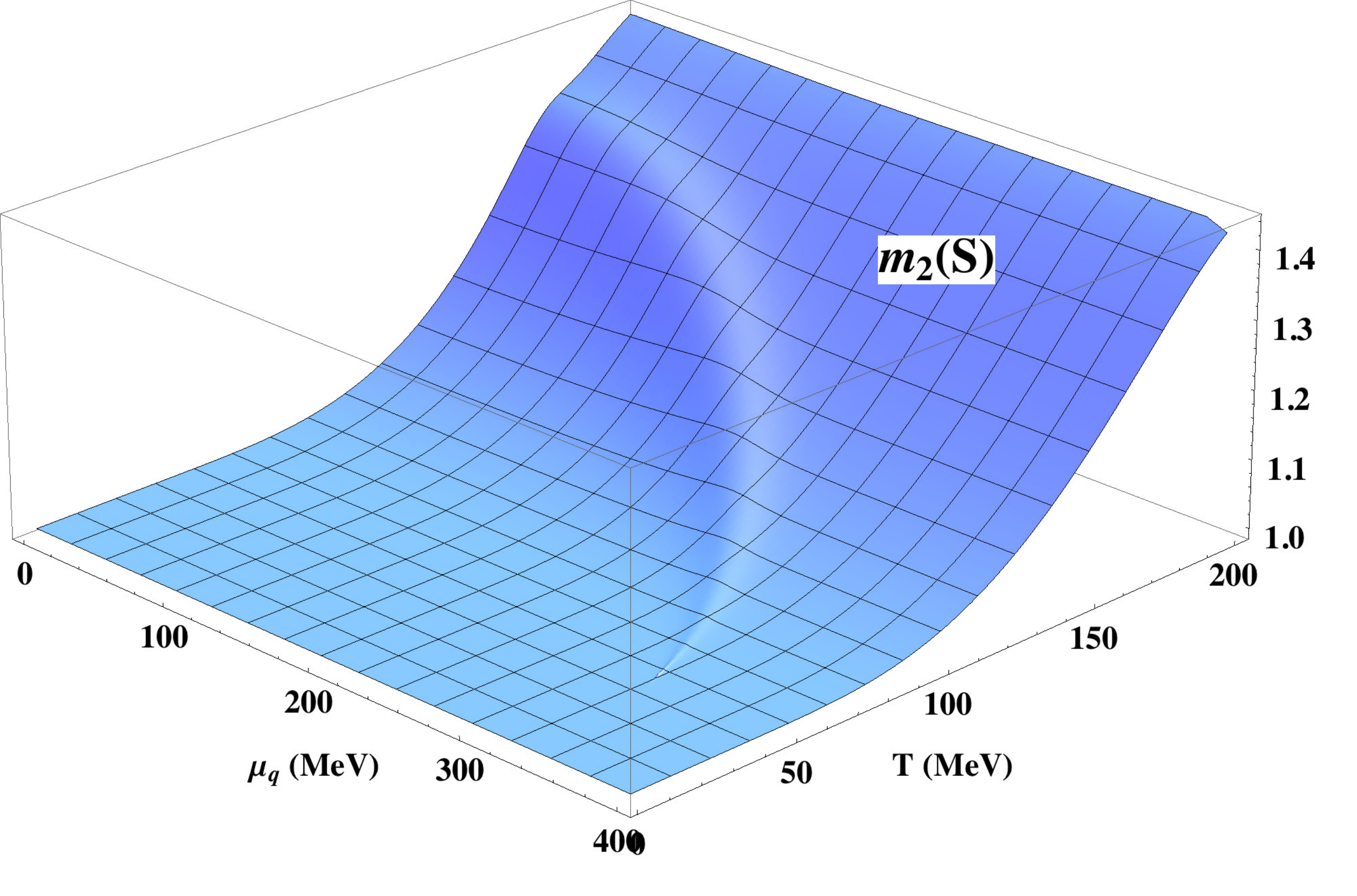}\newline
\includegraphics[width=0.5\textwidth]{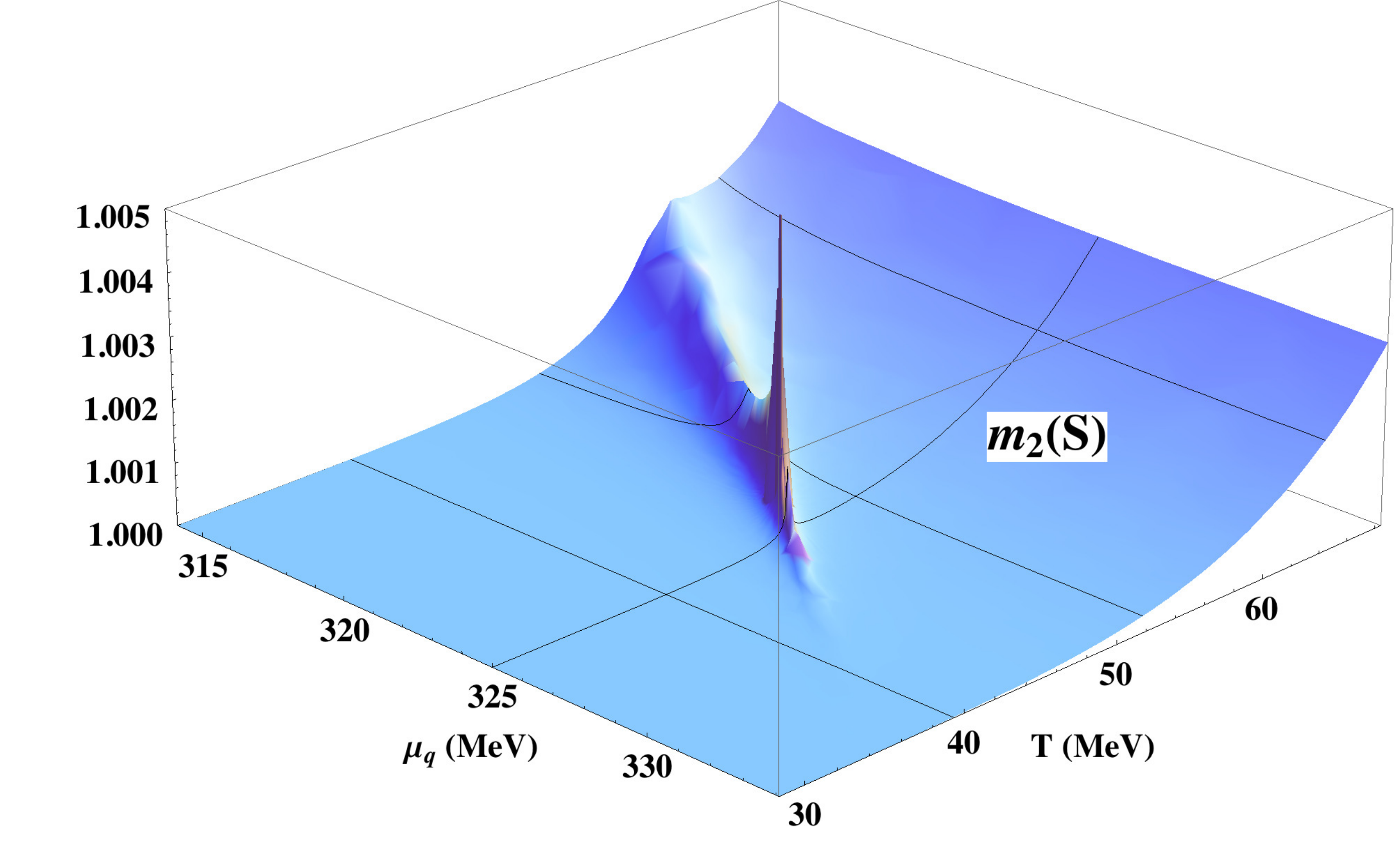} 
\caption{$m_2$ for each of $B,Q,S$ on the phase diagram. }
\label{fig:m2}
\end{figure}

Here we display the numerical results for the fluctuation observables of
primary interest. The volume factors in the susceptibilities are removed by
considering ratios 
\begin{align}
m_1(X)=\frac{T\chi_X^{(3)}}{\chi_{XX}}, \qquad m_2(X)=\frac{%
T^2\chi_{X}^{(4)}}{\chi_{XX}},
\end{align}
for $X=B,Q,S$. We display the results for $m_1,m_2$ for each conserved
charge in Figures \ref{fig:m1} and \ref{fig:m2}. Note that $%
m_1(S)\propto\chi_{sss}=0$, so we do not show it.

\begin{figure}[tbp]
\includegraphics[width=0.4\textwidth]{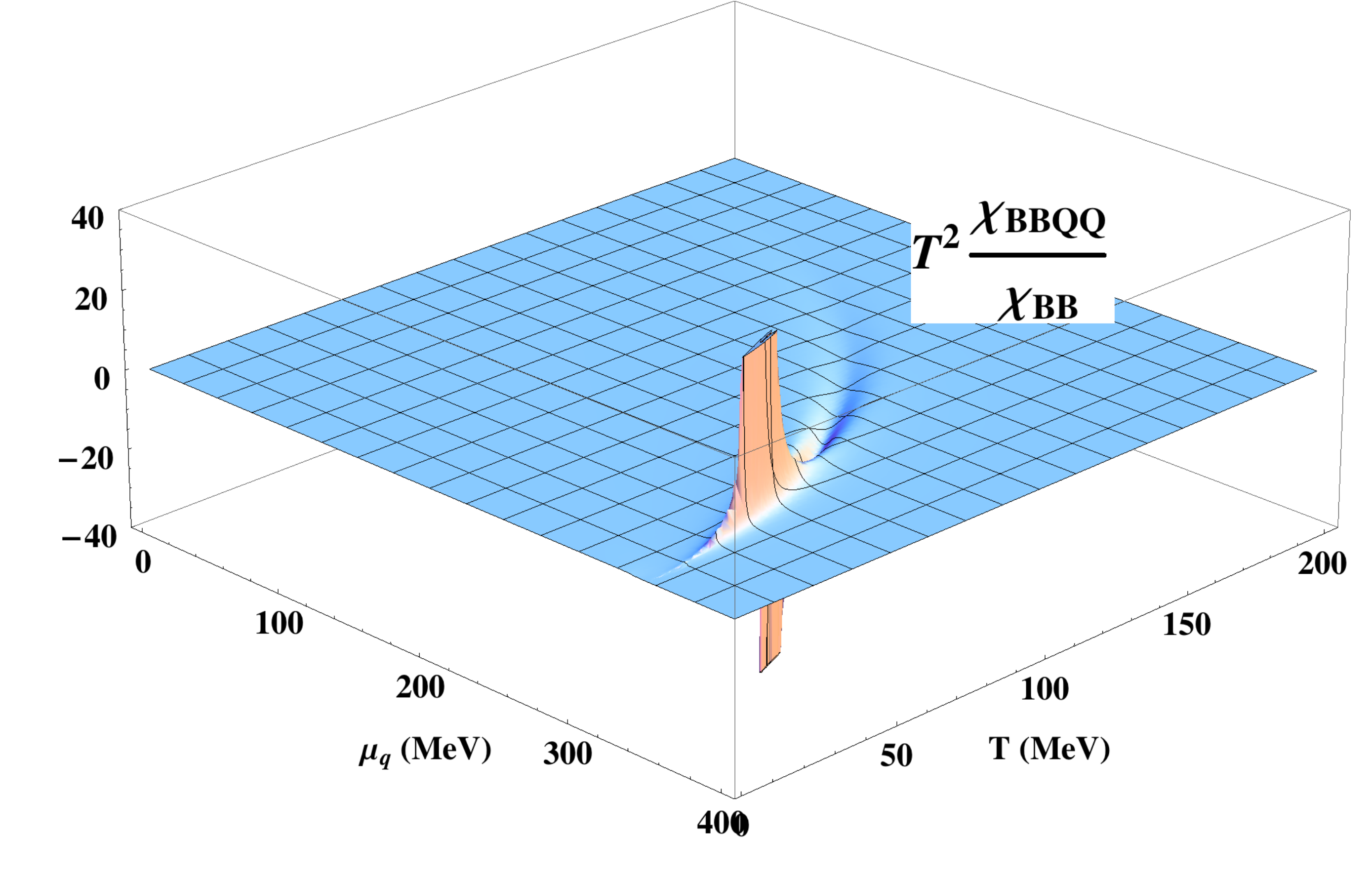}\newline
\includegraphics[width=0.4\textwidth]{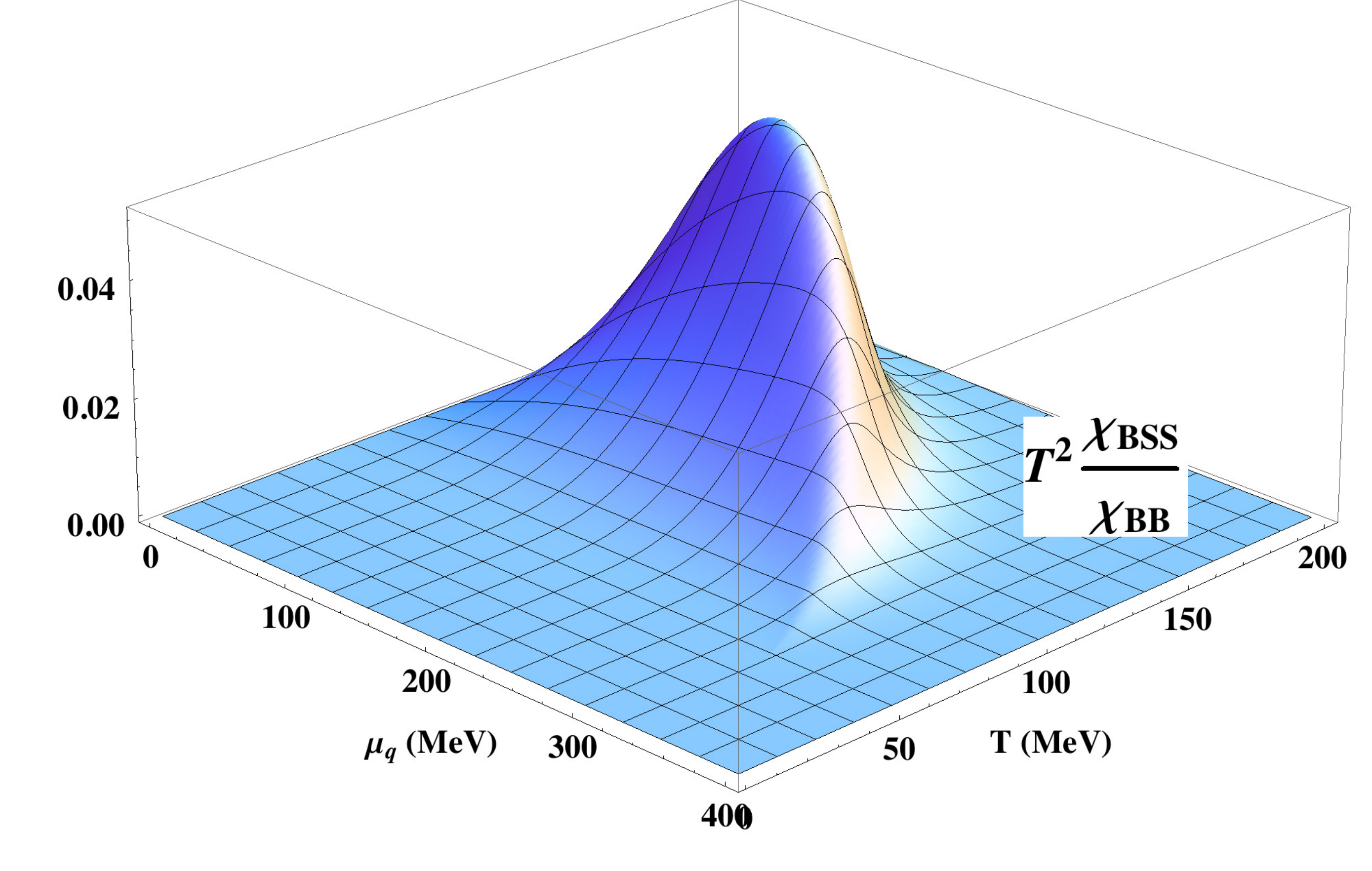}\newline
\includegraphics[width=0.4\textwidth]{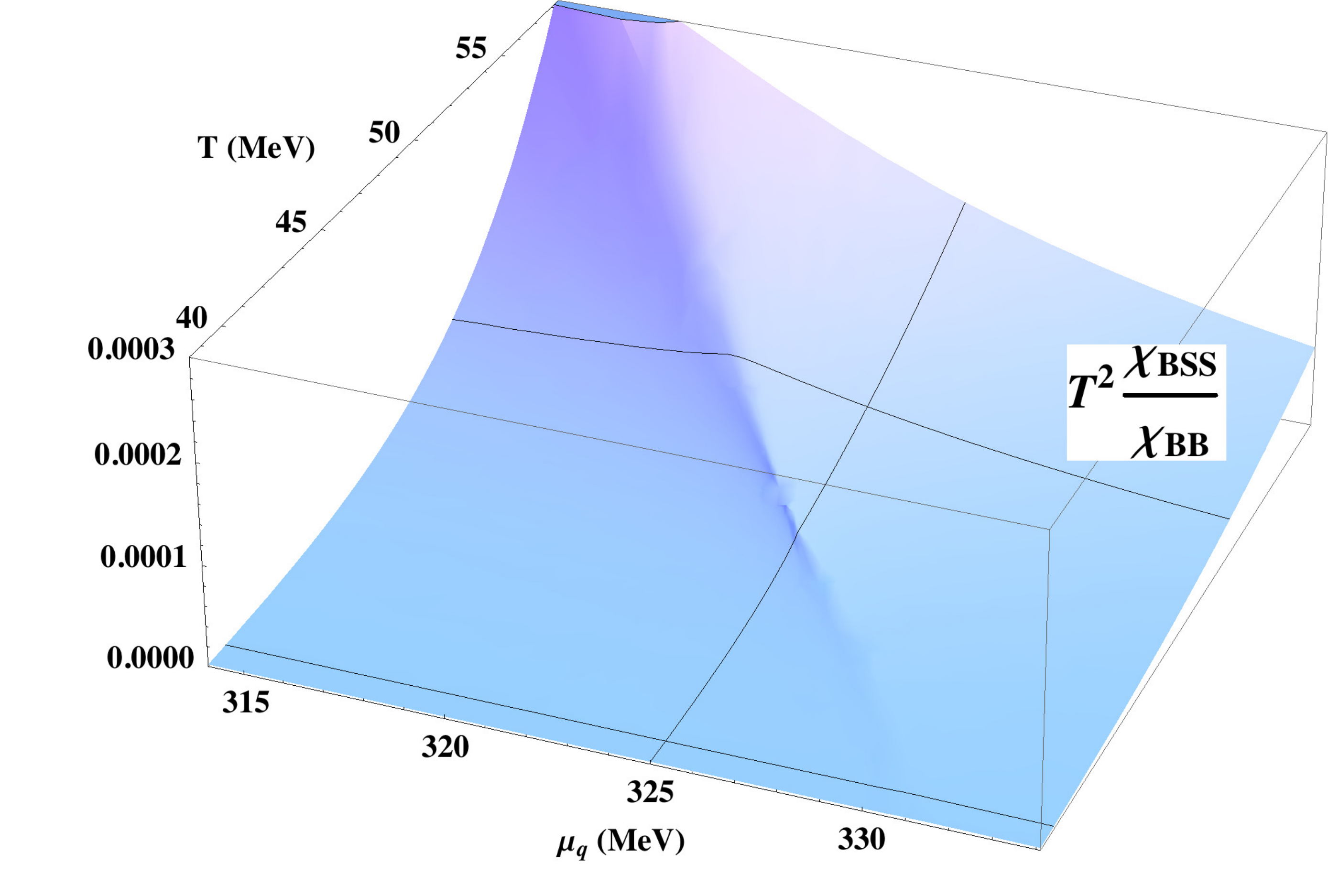} %
\includegraphics[width=0.4\textwidth]{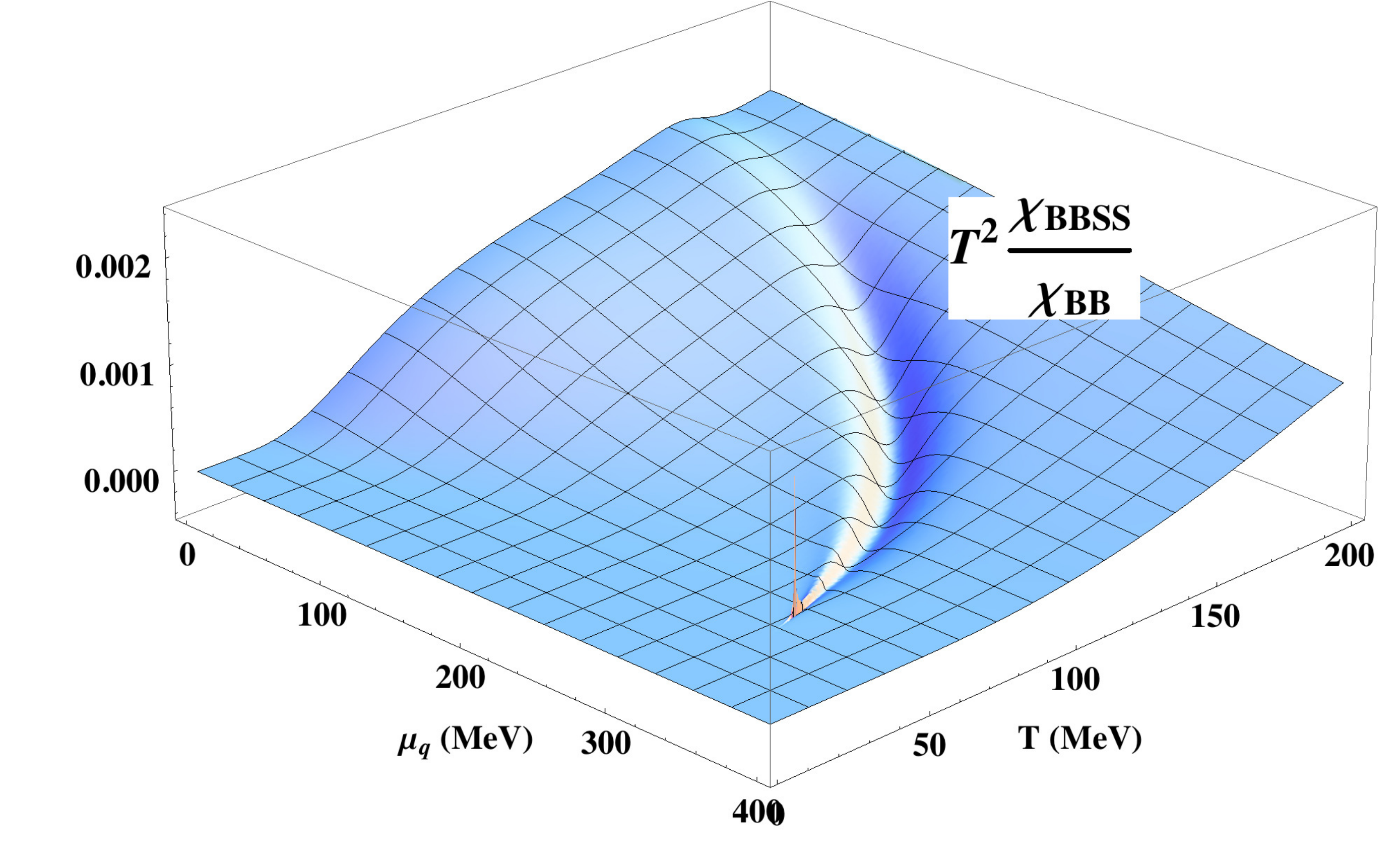}\newline
\includegraphics[width=0.4\textwidth]{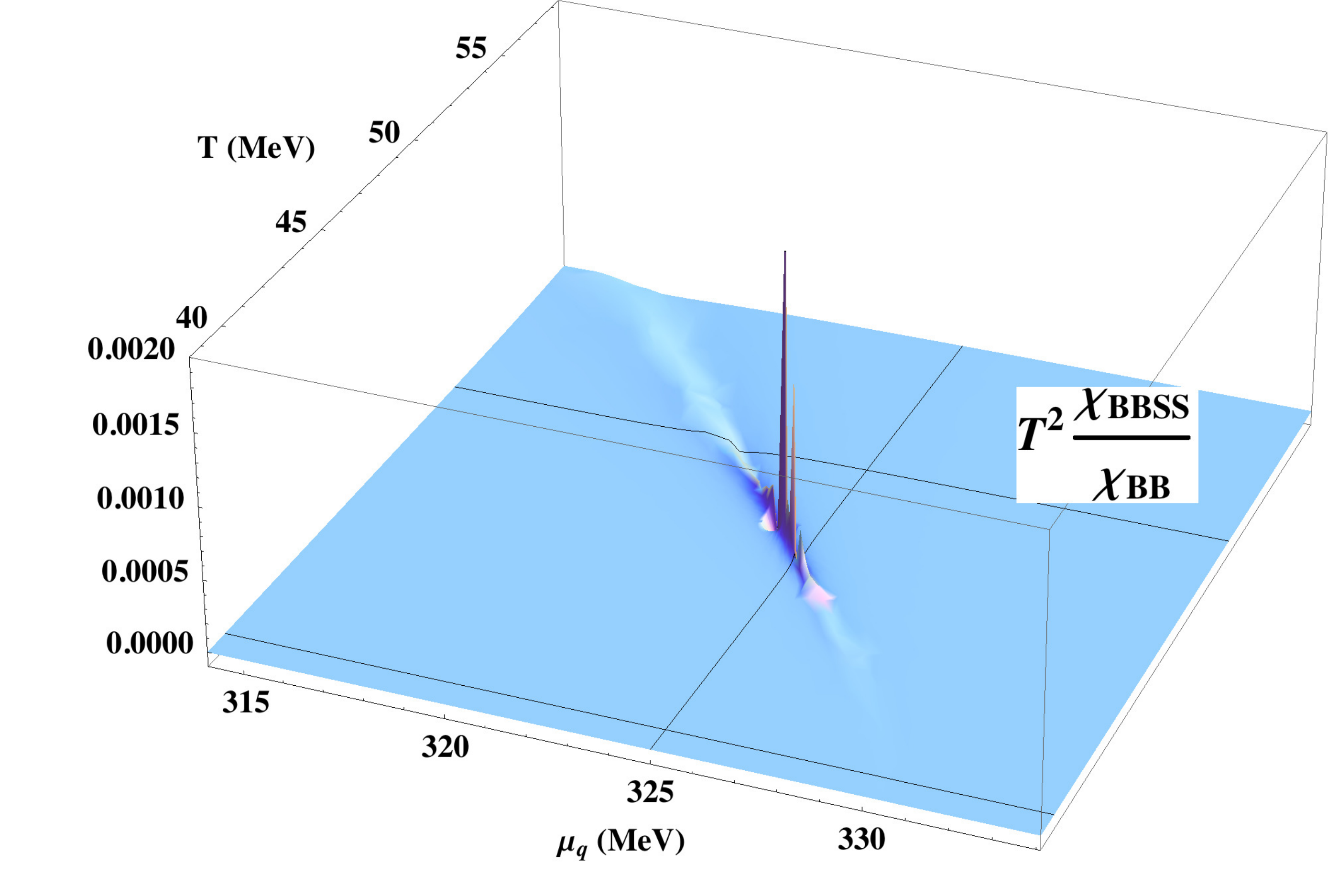}
\caption{Ratios of cross-correlations $T^2\protect\chi_{BBQQ}/\protect\chi_{BB}$, $T^2\protect\chi_{BSS}/\protect\chi_{BB}$, and $T^2\protect\chi_{BBSS}/\protect\chi_{BB}$ on the
phase diagram. }
\label{fig:m2BBSS}
\end{figure}

In $m_{2}(S)$, we see the divergence in $\chi _{s}^{(4)}$ explained above.
However, as seen in the scaling of the axes, the magnitude of the
singularity and area of the critical region are small. It requires high
numerical precision to make the effect visible. Similarly we can form
observable ratios from other third- and fourth-order hadron-level
susceptibilities, involving strangeness or cross-correlations in baryon and charge fluctuations.  The ratio $T\chi _{BSS}/\chi _{BB}$ is proportional to $\chi_{qss}$; however, the divergence in $\chi _{qss}$ is effectively canceled by the divergence in $\chi_{qq}$, as suggested by comparing Fig.\ref{fig:chiqss} and Fig.\ref{fig:suscept2}.  Consequently, the ratio shows no singular behaviour at the CEP. For the ratio $T^{2}\chi_{BBSS}/\chi _{BB}$, $\chi_{uuss}$ in the numerator diverges with a larger power of $\xi$ than the denominator, as seen by comparing the diagrams in Fig.\ref{fig:chi-uuss} and Fig.\ref{fig:suscept2}, so there is a visible singularity peak at CEP. 
The shape of $T^{2}\chi_{BBQQ}/\chi _{BB}$ is very similar to $m_{2}(B)$ and $m_{2}(Q)$, because all three are dominated by the light quark susceptibilities.  The results in Fig. \ref{fig:m2BBSS}
confirm that singularities in the isospin and strangeness susceptibilities
make small impact on the observable fluctuations.

\begin{figure}[tbp]
\hspace*{3.5mm}\includegraphics[width=0.4\textwidth]{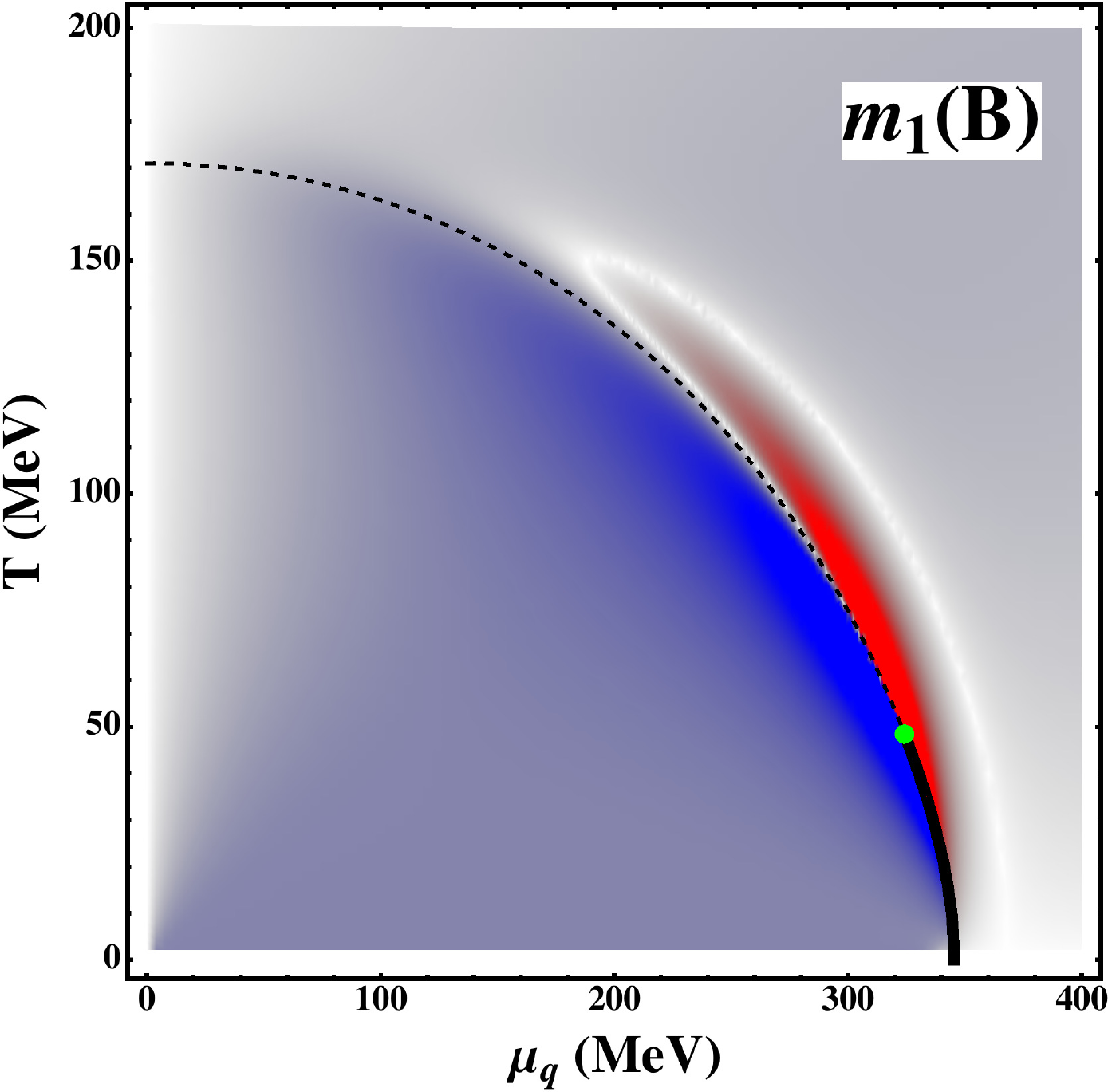}\newline
\includegraphics[width=0.4\textwidth]{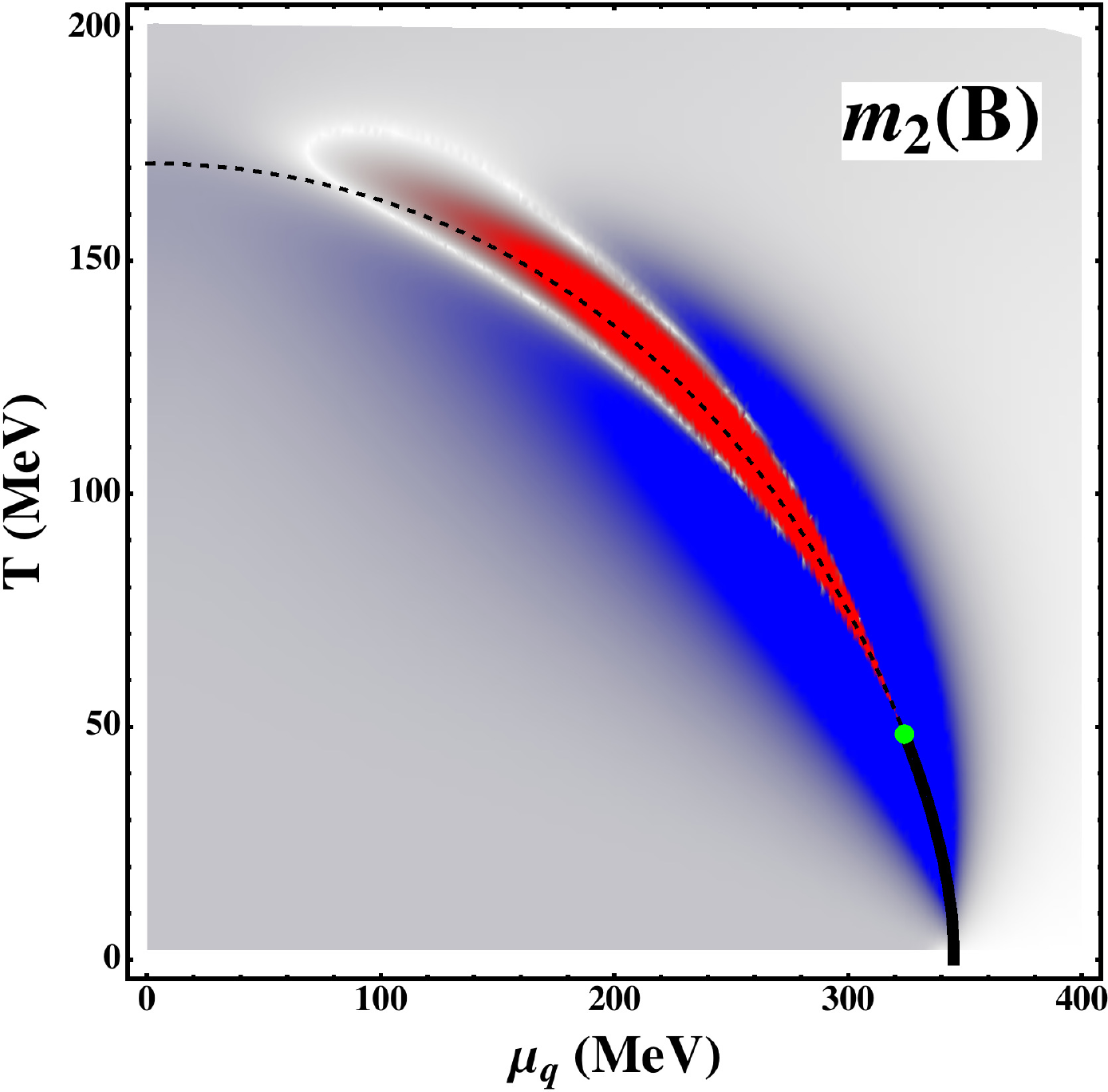}
\caption{For baryon susceptibility, the figures show the sign change of $m_1$
and $m_2$ on the phase diagram. Red areas are negative, and blue are
positive. The green point indicates the location of CEP. The dashed thin
line and the solid thick line stand for the crossover line and the first
order phase transition line respectively. }
\label{fig:sign-change-B}
\end{figure}

The area of the positive and negative regions in the $m_1,m_2$ for $B,Q$ are
seen more clearly in the density plots Figs. \ref{fig:sign-change-B} and \ref%
{fig:sign-change-Q}. The shapes are consistent with previous results \cite%
{Asakawa:2009aj}. Note the area of the critical region for the baryon number
fluctuations is larger than the same region for charge fluctuations,
consistent with the overall larger magnitude of the baryon fluctuations seen
in Figs. \ref{fig:m1} and \ref{fig:m2}. The area of the critical region for
strangeness is smaller still than for the charge fluctuations, and is due to
the small effect of the singularity on the strange quark susceptibilities away
from the critical point.  In the context of the conventional equilibrium 
fluctuations framework to look for signatures of a critical point, the large mass 
of the strange quark (comparable to $T,\mu_q$) suppresses fluctuation observables.
However, this analysis does not account for the possible impact of nonequilibrium 
dynamics, such as critical slowing, on strangeness yields and fluctuations.

\begin{figure}[tbp]
\hspace*{3.5mm}\includegraphics[width=0.4\textwidth]{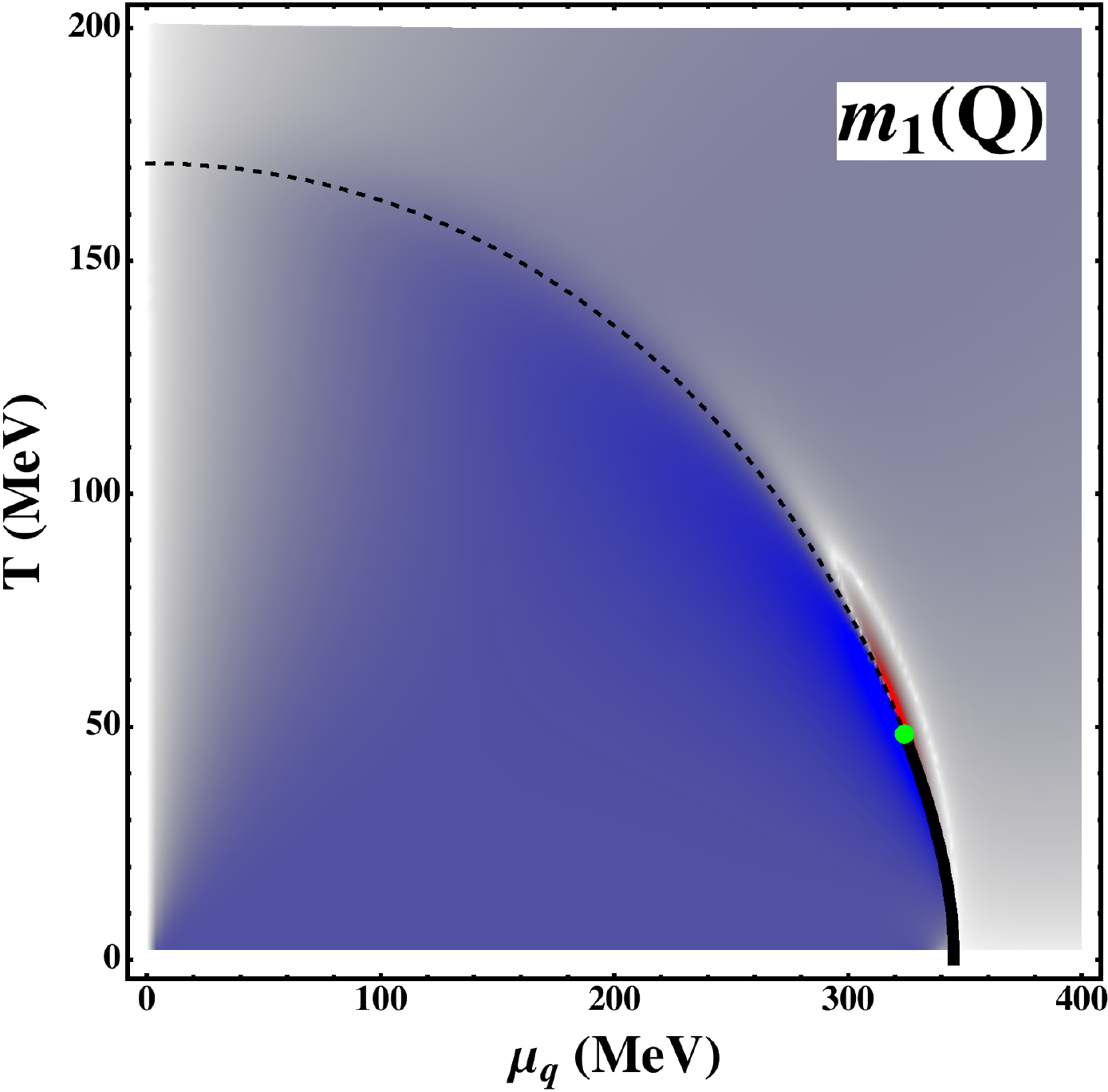}\newline
\includegraphics[width=0.4\textwidth]{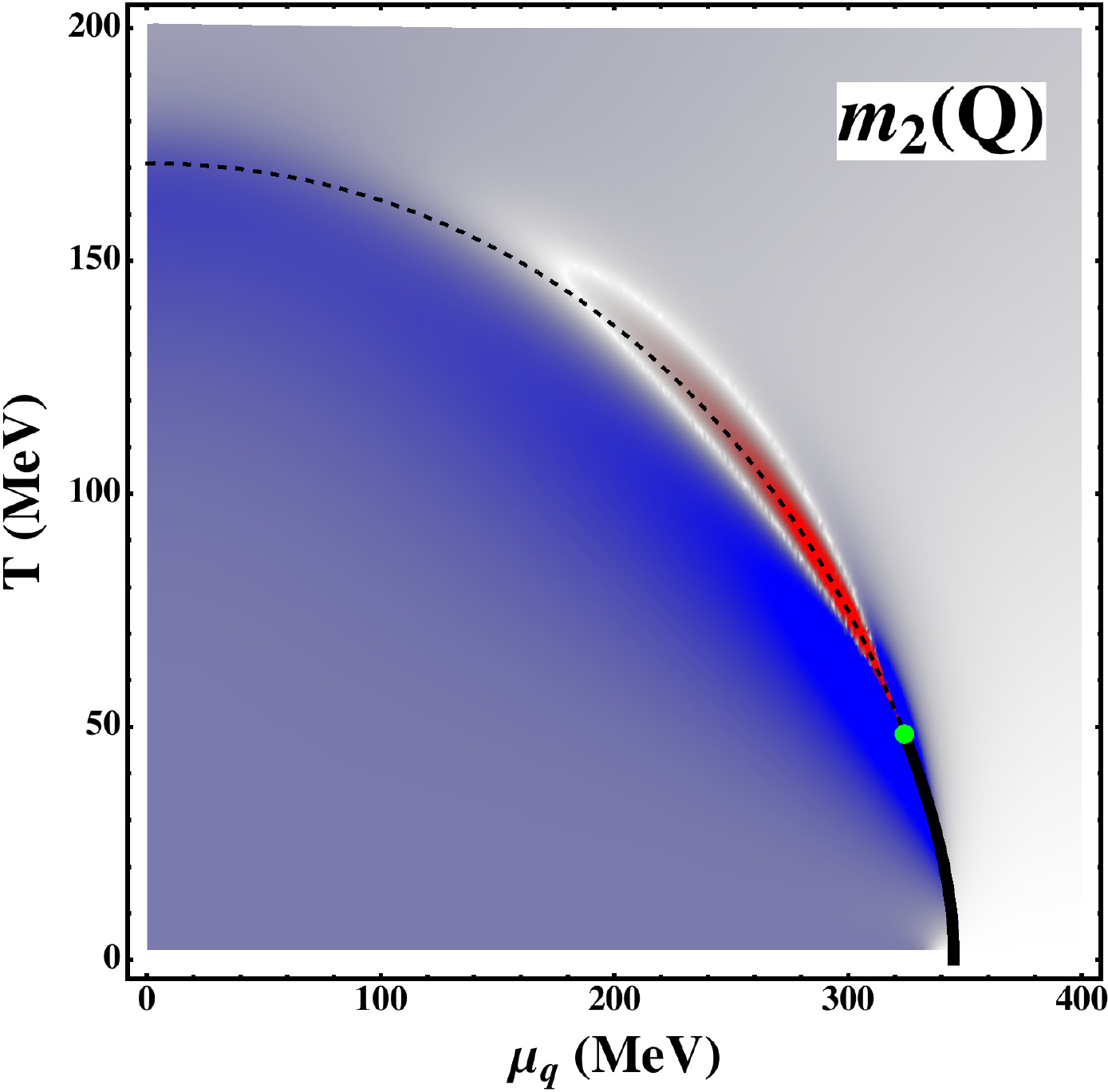}
\caption{Similar to the previous figure, but for electrical charge
susceptibility. }
\label{fig:sign-change-Q}
\end{figure}

\begin{figure}[tbp]
\includegraphics[width=0.4\textwidth]{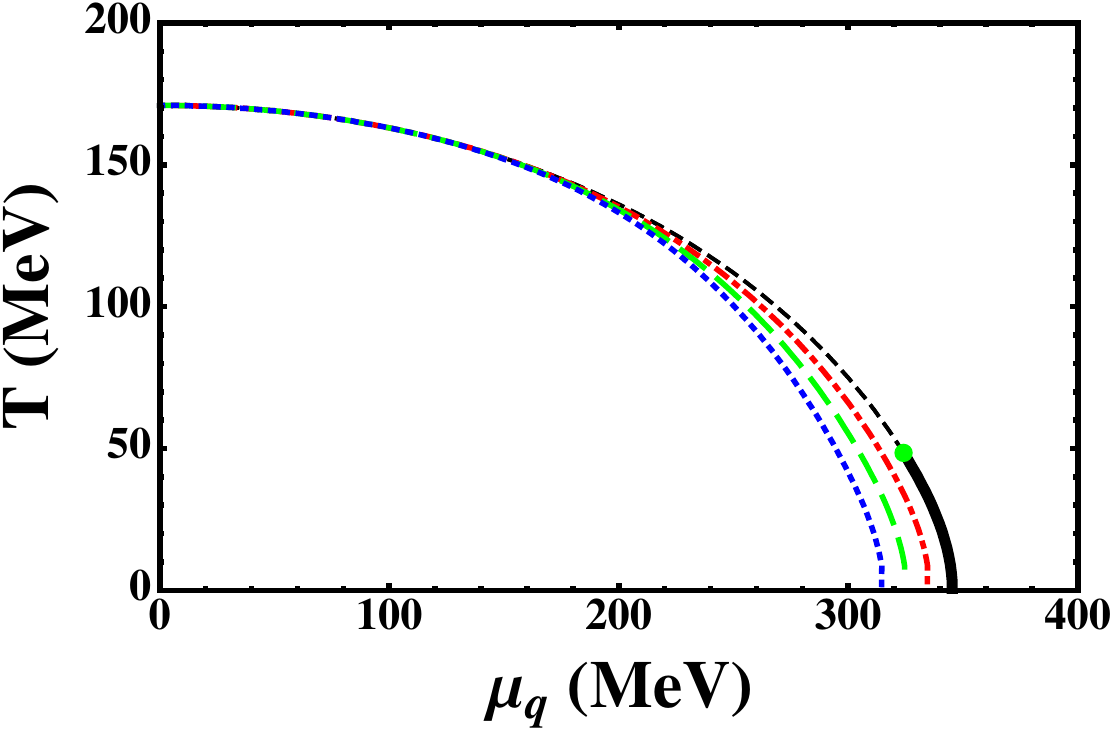}%
\newline
\includegraphics[width=0.4\textwidth]{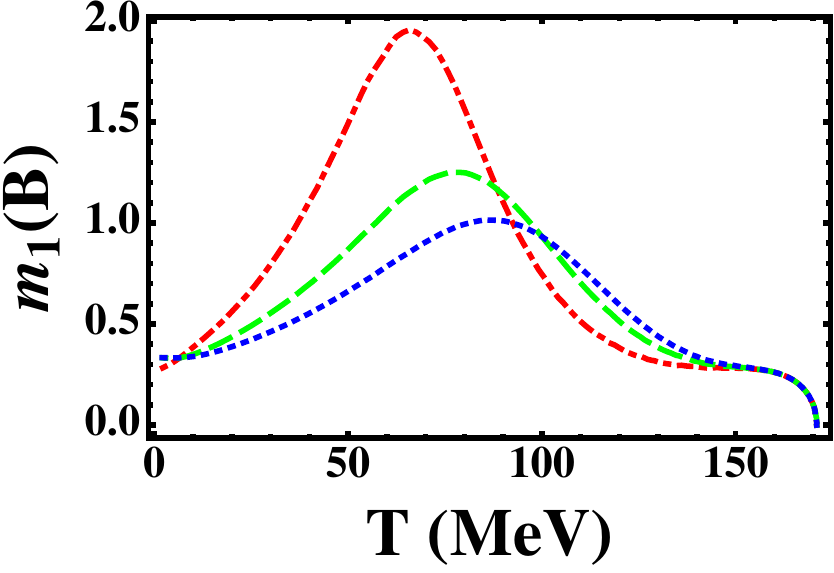}\newline
\includegraphics[width=0.4\textwidth]{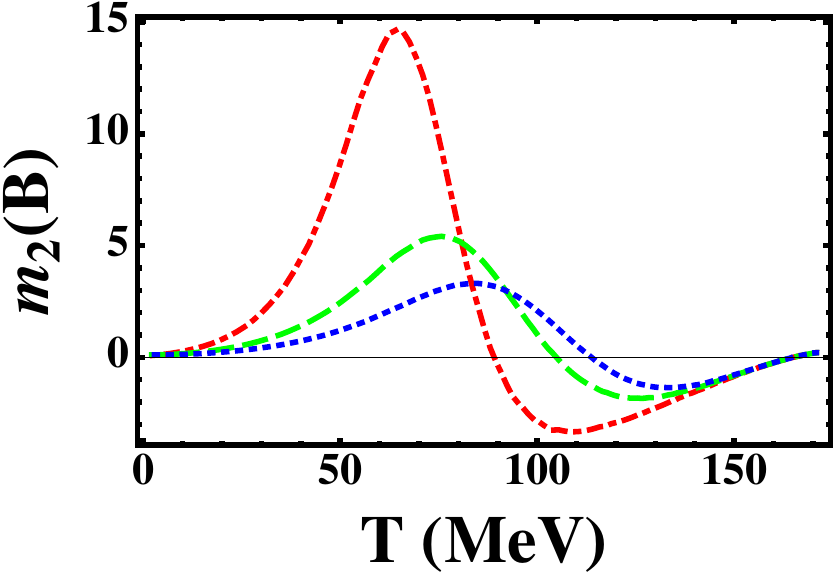}\newline
\caption{Top frame: Hypothetical chemical freeze-out lines on
the phase diagram to model the behaviour of the fluctuation ratios as a function of $\sqrt{s_{NN}}$. Middle and bottom frames:  $m_{1}$ and $m_{2}$ of baryon
number susceptibilities along these freeze-out lines, identified with lines in the top frame by color and dashing. }
\label{fig:B-FO}
\end{figure}

\begin{figure}[tbp]
\includegraphics[width=0.4\textwidth]{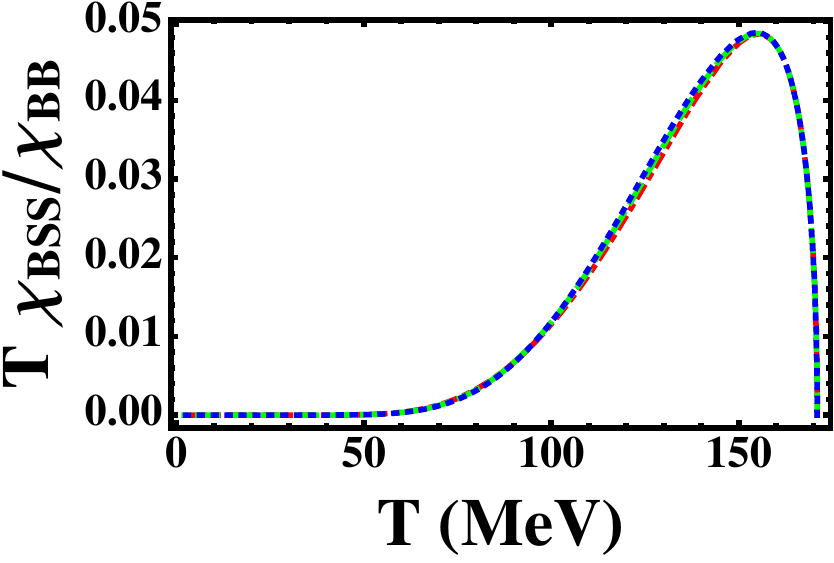}\newline
\includegraphics[width=0.4\textwidth]{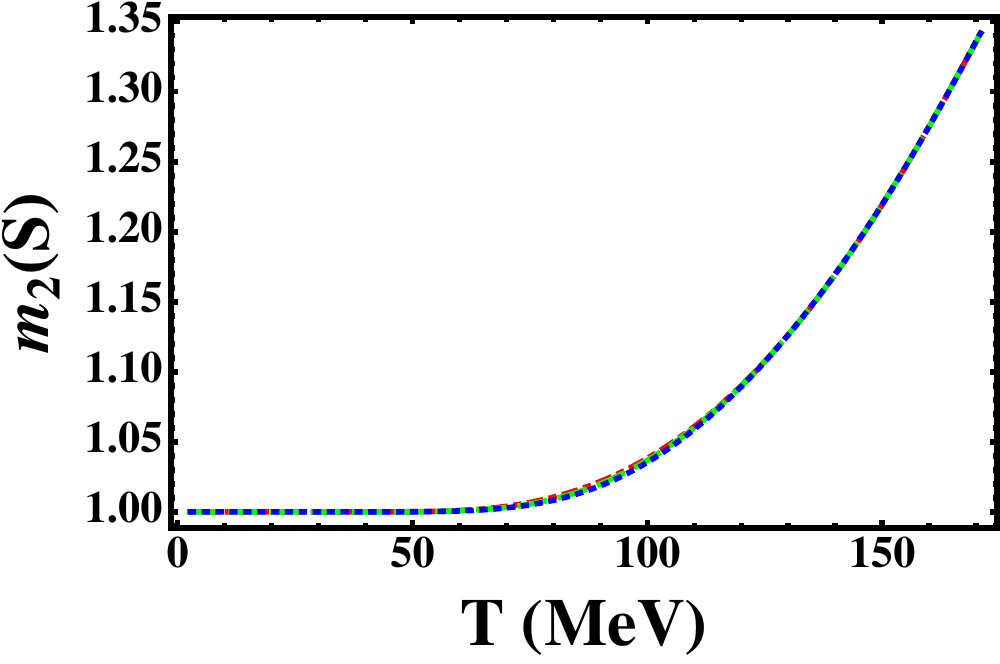}\newline
\includegraphics[width=0.4\textwidth]{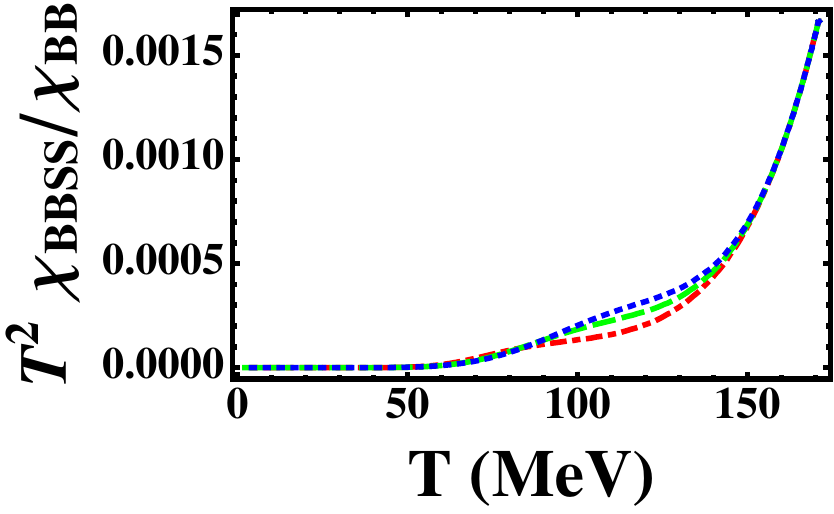}
\caption{The strangeness-dependent ratios $T\protect\chi_{BSS}/\protect\chi_{BB}$, $m_2(S)$ 
and $T^2\protect\chi_{BBSS}/\protect\chi_{BB}$ along the chemical freeze-out
lines  defined in the top frame of Figure \ref{fig:B-FO}. }
\label{fig:S-FO}
\end{figure}

In addition, following \cite{Stephanov:2011pb} and \cite{Chen:2014ufa}, we
show the result of extracting the values of $m_{1}$ and $m_{2}$ along
various hypothetical freeze-out lines chosen to pass varying distances from
the critical point. The freeze-out lines are shown in the top frame of Fig. %
\ref{fig:B-FO}. The behaviour of $m_{1}(B)$ and $m_{2}(B)$ along the
freeze-out lines shows good qualitative agreement with the previous GN model
calculations, supporting the robustness of these shapes and their variation
with distance from the CEP. The profile of charge fluctuations along the
freeze-out line is very similar to the baryon fluctuations, with the
apparent difference due to the overall smaller amplitude of the variation.
In Fig.\ref{fig:S-FO}, we show the ratios involving strangeness along the freeze-out lines, $T^2\protect\chi_{BSS}/\protect\chi_{BB}$, $m_2(S)$ and $T^2\protect\chi_{BBSS}/\protect\chi_{BB}$. As the critical region for strangeness is very small, these ratios are not sensitive to the singularity of the strange quark susceptibilities.

\begin{figure}[tbp]
\includegraphics[width=0.4\textwidth]{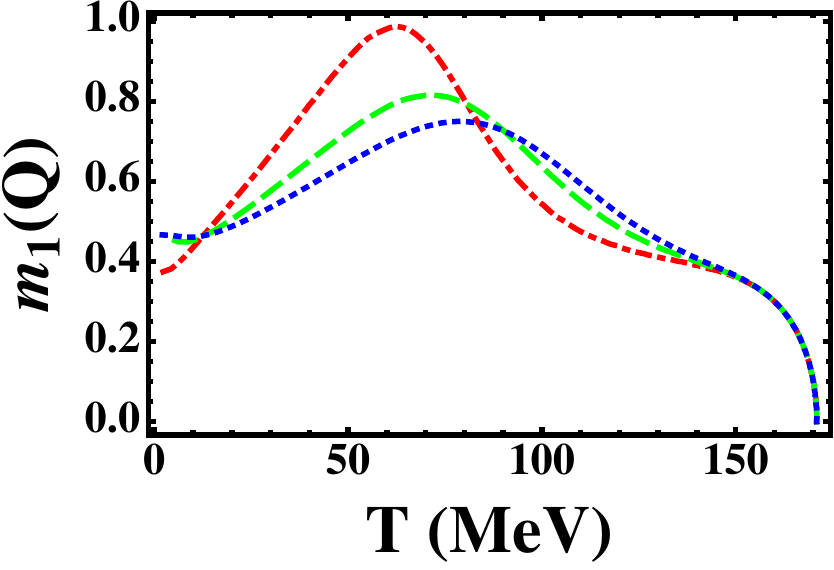}\newline
\includegraphics[width=0.4\textwidth]{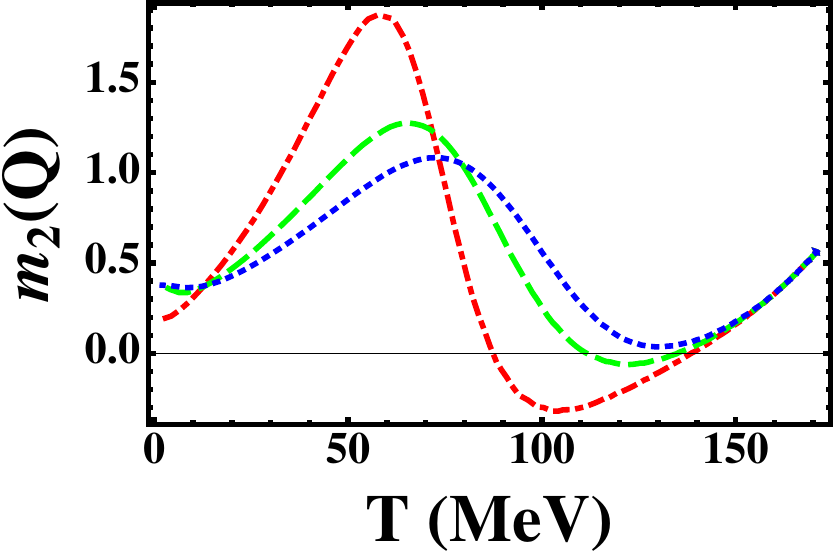}
\caption{Electrical charge $m_1$ and $m_2$ along the freeze-out lines defined in the top frame of Figure \ref{fig:B-FO}.}
\label{fig:Q-FO}
\end{figure}

\begin{figure}[tbp]
\includegraphics[width=0.4\textwidth]{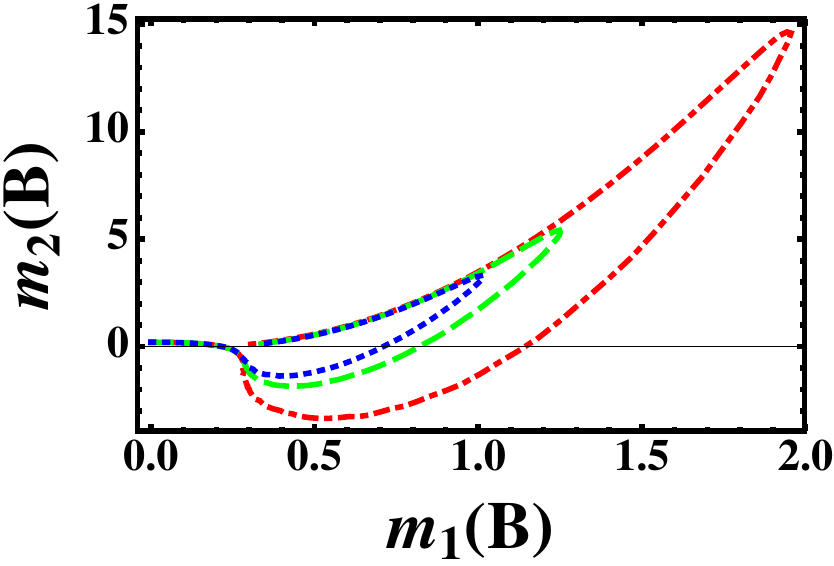}\newline
\includegraphics[width=0.4\textwidth]{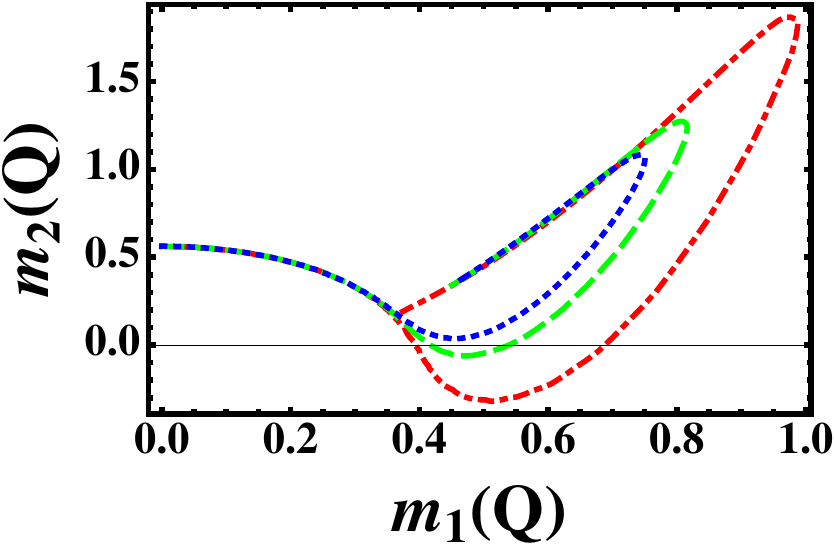}\newline
\includegraphics[width=0.4\textwidth]{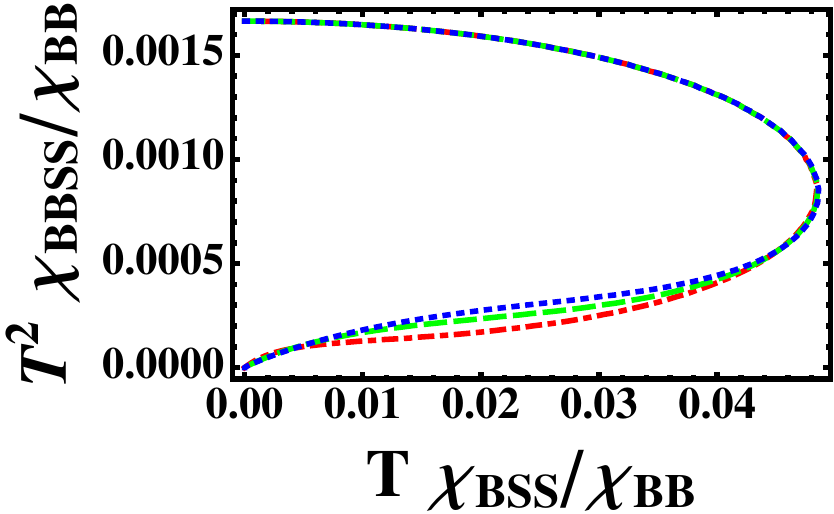}
\caption{Top and middle: $m_2$ plotted against $m_1$ along the chemical freeze-out lines, for both baryon number and electric charge susceptibilities. Bottom: an analogous ratio for the cross-correlation between baryon number and strangeness.  Color and dashing of the line identifies it with the freeze-out lines defined in Fig.\,\ref{fig:B-FO}. }
\label{fig:m1-m2-B-Q-FO}
\end{figure}

Finally, we show how $m_{1}$ and $m_{2}$ vary in relation to each other
along the freeze-out line.\footnote{We thank R. Gavai for suggesting this plot.}
 In Fig. \ref{fig:m1-m2-B-Q-FO}, decreasing $\sqrt{%
s_{NN}}$ (increasing $\mu _{B}$ or decreasing $T$) starts from $m_{1}=0$
and continues in an anti-clockwise trajectory around the loop. The
convergence of the lines at high $T$ (close to $m_{1}=0$) is due to the
convergence of the hypothetical freeze-out lines in our modeling, see Fig. %
\ref{fig:B-FO} (top). For the lower $T$ freeze-out lines (green dashed and
blue dotted lines), the loop does not close. Even so, the low $T$/high $\mu $
end of the trajectory is the same for all freeze-out lines because the
statistics and fluctuations are given by $m,\mu $-dominated thermal
distributions there. These plots show clearly the ordering of features 
\begin{equation}
T_{\mathrm{min},m_{2}}>T_{\mathrm{max},m_{1}}>T_{\mathrm{max},m_{2}}>T_{%
\mathrm{CEP}}  \label{Tordering}
\end{equation}%
which was found also in the GN and Ising models.  (In the Ising model, we 
plot the kurtosis of the magnetization $\partial^4M/\partial H^4$ versus
the skewness $\partial^3M/\partial H^3$ along lines of constant $H$, yielding
a plot of the same shape as Fig. \ref{fig:m1-m2-B-Q-FO}.)
Again, the magnitudes of the $m_{1},m_{2}$ are much greater for
baryon fluctuations than for charge fluctuations. We expect these
characteristics are robust. Nonequilibrium effects and higher-order,
model-dependent corrections will perturb the loop. However, given also the
similarity to the GN and Ising results, it appears large corrections would
be necessary to affect the ordering in Eq.\thinspace (\ref{Tordering}). For ratios with strangeness, this ordering is not seen, because the critical region for strangeness is too small to impact the observables along the chosen freeze-out lines.

\section{Conclusion}

In this paper, we have investigated the behaviour of non-Gaussian
fluctuations of baryon number, electric charge and strangeness on the phase
diagram. Our purpose was to characterize their dependence at $\mu \neq 0$
and find robust qualitative features that can be used to help interpret
experimental data in the search for a QCD critical point. To this end, we
evaluated the third- and fourth-order susceptibilities in the NJL model,
which is in the same universality class as QCD. Our analysis is limited to
the mean-field approximation which means that we have not included several
known corrections: modifications to the critical exponents controlling the
divergence of the correlation length at the CEP, and model-dependent
corrections described by higher-dimensional operators in the low energy
effective field theory. The corrections to the critical exponents are better
known quantitatively, but model-dependent corrections are less
well-controlled and their importance is not known in the context of the
physical fireball where the correlation length of the critical mode may not
be much larger than thermal correlation length.

We have found several characteristics of the third- and fourth-order
susceptibilities that are likely to be robust. These are:

\begin{enumerate}
\item Baryon number fluctuations are the largest in amplitude. This is
because they are dominantly composed of the leading singular contributions
from the light quark susceptibilities. Charge fluctuations are suppressed by
numerical coefficients, as seen by comparing the expansions in \req{3susceptibilities} and \req{4susceptibilites}. Strangeness fluctuations are the
smallest in amplitude, though at high order they also show a singularity at
the critical point. The strangeness fluctuations are suppressed by the
relatively large bare mass of the strange quark, $m_{s}\sim T$.

\item The \textquotedblleft critical region\textquotedblright\ where we see
non-monotonicity in $m_{1}$ and $m_{2}$ along hypothetical freeze-out lines%
is largest for baryon number fluctuations. Therefore it provides
the largest signal for freeze-outs occurring farther in the $\mu ,T$ plane
from the CP. This act follows from the amplitude of the fluctuations being
largest.

\item The qualitative profile of the third and fourth order susceptibilities
along the freeze-out line is model-independent. $m_{1}$ for both $B,Q$ has a
single peak at a temperature greater than the CP. $m_{2}$ has first a
minimum and then a peak as the critical point is approached from high $T$
(equivalently larger collision energy $\sqrt{s_{NN}}$). The $m_{2}<0$ negative
region may be accessible at chemical freeze-out in the symmetry broken
phase. These behaviours have been seen in the GN and NJL models.

\item The features of the $m_{1}$ and $m_{2}$ freeze-out profiles obey a
numerical ordering temperature or collision energy, given by Eq.\thinspace (%
\ref{Tordering}). This is demonstrated by plotting $m_{2}$ versus $m_{1}$
along the freeze-out line, and we have argued this trajectory is difficult
to change in qualitative features.
\end{enumerate}

It is our hope that these features can aid the interpretation of
experimental data for the third- and fourth-order susceptibilities. Although
our results may support preliminary indications of critical behaviour in the
data, it is clear that much more work is required to identify these
signatures with a QCD critical point.

\begin{appendix} 
\section{NJL model}

In our study, we consider the 3-flavor NJL model with two degenerate light quarks and a heavier strange quark.  The quark level Lagrangian is
\begin{align}
\mathcal{L}= \:&
\sum_{i=u,d,s}\bar \psi_i(\slashed{\partial}-m_i)\psi_i 
+ G\Tr[(\bar\psi\lambda_a\psi)^2 +(\bar\psi\gamma_5\lambda_a\psi)^2] \notag \\
& -K(\det\bar q(1-\gamma_5)q+\det\bar q(1+\gamma_5)q)
\end{align}
The model parameters are the ``bare'' quark masses $m_i$ and the 4-fermion couplings $G,K$.  In addition, for numerical evaluation of the effective potential below we must introduce a momentum cutoff $\Lambda$.  These model parameters are fixed by matching the pion, kaon, $\eta'$ masses and the pion decay constant in vacuum, $m_\pi=138$\,MeV, $m_K=495$\,MeV, $m_{\eta'}=958$\,MeV and $f_\pi=93$\,MeV.  We use the residual freedom to set the light quark mass near the physical point, so that the parameter set is
\begin{align}
m_u=m_d=5\,\mathrm{MeV},\quad 
m_s=136\,\mathrm{MeV}, \\
\Lambda=631\,\mathrm{MeV},\quad
G\Lambda^2=1.83, \quad
K\Lambda^5=9.29
\end{align}
Spontaneous symmetry breaking leading to nonzero light quark condensates $\langle\bar u u\rangle,\langle\bar dd\rangle\neq 0$ has greater impact on phenomenology than the explicit breaking from their masses, as is the case in QCD.

To solve the model and study the phase diagram, we take the mean field approximation.  After shifting the quark bilinears with their vacuum expectation values $\bar q q\to \langle\bar q q\rangle+\varphi_q$, the quark fields can be integrated out, and we obtain an effective potential depending only on $T,\mu_q$ and $\varphi_q$ ($q=u,d,s$).  As long as the isospin chemical potential
\begin{align}\label{onecondensate}
\mu_I=\mu_u-\mu_d \lesssim \frac{m_\pi}{2}\simeq 65 ~\mathrm{MeV}
\end{align}
the $u,d$ condensates are equal and transition at the same $\mu,T$.  Pseudoscalar, vector and pseudovector diquark condensates are also suppressed and can be ignored.  Experimental data indicates that this condition holds even at the lowest collision energies.  The resulting effective potential has the form
\begin{align}
&\Omega=2G(\sigma_u^2+\sigma_d^2+\sigma_s^2)-4K\sigma_u\sigma_d\sigma_s+\sum_{f=u,d,s}\Omega_f\\
&\Omega_f=-2N_c\int\frac{d^3p}{(2\pi)^3}\Big(E_f\Theta(\Lambda^2-p^2) 
 \\ \notag &~~~~
 +T\ln(1+e^{-\beta(E_f-\mu_f)})+T\ln(1+e^{-\beta(E_f+\mu_f)}\Big)
\end{align}
where $\Theta(x)$ is the Heaviside step function, $E_f^2=\vec p^2+M_f^2$.  The last two terms in $\Omega_f$ are respectively the particle and antiparticle fermi distributions including the chemical potential.  The $M_f$ are effective masses, functions of the condensates
\begin{align}
M_f=m_f-4G\sigma_f+2K\sigma_{f'}\sigma_{f''},   \quad  (f\neq f'\neq f'')
\end{align}
The term off-diagonal in flavor implies that the strange quark condensate is discontinuous where the light quark condensates are discontinuous and vice versa.  The larger bare mass of the strange quark means the impact of explicit chiral symmetry breaking is larger than for the light quarks, and its phase transition occurs only for larger chemical potential and temperature, see \cite{Chen:2014ufa} for discussion how the position of the critical point depends on the bare quark mass.  If a critical point is accessible to lattice or HIC, it will be the light quark critical line, which is at lower chemical potential and temperature.  However, due to the flavor coupling, some evidence of the criticality will be manifest in the strange susceptibilities.  We will discuss this point further below.

The phase diagram is determined by solving the coupled set of gap equations
\begin{align}
0  = \frac{\partial\Omega}{\partial\sigma_\alpha} = \:&\: \sum_\gamma \left(4G\delta_{\alpha\gamma}-2K\sum_\theta|\epsilon_{\alpha\gamma\theta}|\sigma_\theta\right) \Delta_\gamma  \\
\Delta_\gamma = \:&\: \sigma_\gamma-\frac{\partial\Omega_\gamma}{\partial m_\gamma} \notag
\end{align}
taking the solution that corresponds to the global minimum of the effective potential.  ($|\epsilon_{\alpha\gamma\theta}|$ tensor is +1 in entries with $\alpha\neq\gamma\neq\theta$.)  We solve the system using two independent numerical methods, as a quantitative check on our results.

\section{Diagrammatic system}\label{app:diagrams}

$\mu$-derivatives of the pressure include a term for each flavor
\begin{align}
\frac{dP(T,\vec{\mu},\vec{\sigma})}{d\mu_\alpha}=\frac{\partial P}{\partial \mu_\alpha}+\sum_\beta \left.\frac{\partial P}{\partial\sigma_\beta}\right|_{\vec{\sigma}^{(0)}}\frac{\partial\sigma_\beta}{\partial\mu_\alpha}
\end{align}
The vectors $\vec{\mu},\vec{\sigma}$ are shorthand for $(\mu_u,\mu_d,\mu_s),(\sigma_u,\sigma_d,\sigma_s), $ and $\vec{\sigma}^{(0)}$ subscript indicates the derivative is evaluated at the minimum of the potential.   The last factor $\partial\sigma_\beta/\partial\mu_\alpha$ can be rewritten using the fact that the gap equation is independent of the chemical potential
\begin{align}
0\:&=\frac{d}{d\mu_\alpha}\left(\left.\frac{\partial P}{\partial\sigma_\beta}\right|_{\vec{\sigma}^{(0)}}\right) \\ \notag
 \:&=\left.\frac{\partial^2P}{\partial\mu_\alpha\partial\sigma_{\beta}}\right|_{\vec{\sigma}^{(0)}}+\sum_{\gamma}\left.\frac{\partial^2P}{\partial\sigma_\beta\partial\sigma_\gamma}\right|_{\vec{\sigma}^{(0)}}\frac{\partial\sigma_\gamma}{\partial\mu_\alpha}
\end{align}
Note that the second term contains a factor which has the form of a two-point correlator of the $\sigma$ field.

When we write out a general (second order) susceptibility, we find two terms
\begin{align}\label{chi2}
\chi_{\alpha\beta}=\:&\frac{d^2 P}{d\mu_id\mu_j} \notag \\
=\:& \frac{\partial^2 P}{\partial\mu_\alpha\partial\mu_\beta}-\sum_{\gamma,\delta} \frac{\partial^2P}{\partial\mu_\alpha\partial\sigma_\gamma}\left(\left.\frac{\partial^2P}{\partial\sigma_\gamma\partial\sigma_\delta}\right|_{\vec{\sigma}^{(0)}}\right)^{-1}\frac{\partial^2P}{\partial\sigma_\delta\partial\mu_\beta}
\end{align}
A similar relation has been discussed by \cite{Fujii:2003bz}. The first is the second derivative at the minimum of the potential, and the second relates $\mu$-derivatives at different points through the $\sigma$ two-point function.  We express this naturally by the diagrams in Figure \ref{fig:suscept2}.  This ``$\sigma$ correlator'' 
\begin{align}\label{sigmacorrelator}
\frac{\partial^2 \Omega}{\partial\sigma_\alpha\partial\sigma_\beta}= \:&\:
 \sum_\gamma \left(4G\delta_{\alpha\gamma}-2K\sum_\theta(\epsilon_{\alpha\gamma\theta})^2\sigma_\theta\right) \frac{\partial\Delta_\gamma}{\partial\sigma_\beta}  \\
\frac{\partial\Delta_\gamma}{\partial\sigma_\beta} = \:&\: \delta_{\gamma\beta}-\frac{\partial^2\Omega_\gamma}{\partial m_\gamma^2} \left(4G\delta_{\beta\gamma}-2K\sum_\theta|\epsilon_{\beta\gamma\theta}|\sigma_\theta\right)  \notag
\end{align} 
is not diagonal in flavor space due to the anomaly-induced interaction (proportional to $K$).

The third and fourth order susceptibilities include new terms, such as the $\sigma$ three- and four-point functions.  By writing out all the diagrams constructed from these pieces, one may search higher order susceptibilities for their most singular contributions.    

\end{appendix}

\vskip0.2cm 
\textit{Acknowledgments}: We would like to thank R. Gavai  and C. Markert for suggestions and discussions.
J.D. is supported in part by the Major State Basic
Research Development Program in China (Contract No. 2014CB845406), National
Natural Science Foundation of China (Projects No. 11105082). J.W.C. is
supported in part by the MOST, NTU-CTS, NTU-CASTS of R.O.C., and the DFG and
NSFC (CRC 110). H.K. is supported by Ministry of Science and Technology
(Taiwan, ROC), through Grant No. MOST 103-2811-M-002-087. L.L. is supported
by NNSA cooperative agreement de-na0002008, the Defense Advanced Research
Projects Agencys PULSE program (12-63-PULSE-FP014), the Air Force Office of
Scientific Research (FA9550-14-1-0045) and the National Institute of Health
SBIR 1 LPT\_001.


\end{document}